\documentclass[submission]{SciPost}
\pdfoutput=1

\usepackage{float}
\restylefloat{table}
\usepackage{microtype}
\usepackage{xcolor}
\usepackage[most]{tcolorbox}
\usepackage[utf8]{inputenc}
\usepackage{amsmath,amssymb,amsfonts,amsthm,mathtools,mathrsfs,slashed}
\usepackage[sort&compress,numbers]{natbib}
\usepackage{tocloft}

\definecolor{yellow1}{HTML}{ffffcc}
\definecolor{myviolet}{HTML}{007B8B}
\definecolor{mylavender}{HTML}{DEFEFF}
\definecolor{soft}{HTML}{fff6ed}
\definecolor{sky}{HTML}{CEFFFF}
\definecolor{indigo}{HTML}{006080}
\definecolor{myblue}{HTML}{056FB1}
\definecolor{maroon}{HTML}{db0000}
\definecolor{forest}{HTML}{06961c}
\definecolor{lime}{HTML}{fcffad}
\definecolor{mygreen}{HTML}{024f16}
\definecolor{newgreen}{HTML}{aaf50a}
\definecolor{ocre}{HTML}{eb6315}

\usepackage{hyperref}
\usepackage{prettyref}
\newcommand{\pref}{\prettyref}
\DeclareMathOperator{\tr}{tr}
\DeclareMathOperator{\str}{str}

\hypersetup{
	colorlinks=true,
	linkcolor=indigo,
	filecolor=cyan,      
	citecolor=ocre,
    urlcolor=myblue
}
\begin{document}
	\begin{samepage}
		\begin{flushleft}\huge{\textbf{One loop in $D=11$ vs $D=10$: 4-point check}}\end{flushleft}
		\vspace{20pt}
		{\color{myviolet}\hrule height 1mm}
		
		\vspace*{10pt}
		\begin{flushleft}
		\large 	\textbf{Aviral Aggarwal}$\,{}^a$, \textbf{Subhroneel Chakrabarti}$\,{}^a$, \textbf{Steven Weilong Hsia}$\,{}^a$,\\ \textbf{Ahmed Rakin Kamal}$\,{}^{a\,,\,b}$, \textbf{Linus Wulff}$\,{}^a$
		\end{flushleft}

		\begin{flushleft}
			\emph{\large ${}^a$ Department of Theoretical Physics and Astrophysics,\\ Faculty of Science, Masaryk University, 611 37 Brno, Czech Republic.} \\
               \emph{\large ${}^b$  Department of Mathematics and Natural Sciences,\\ BRAC University, 66 Mohakhali, Dhaka 1212, Bangladesh.}
			\\ \vspace{1mm}
            \href{mailto:aviral@mail.muni.cz}{aviral@mail.muni.cz},
            \href{mailto:hsiasteven000@gmail.com}{hsiasteven000@gmail.com},
            \href{mailto:subhroneelc@physics.muni.cz}{subhroneelc@physics.muni.cz},
            \href{mailto:ahmedrakinkamaltunok@gmail.com}{ahmedrakinkamaltunok@gmail.com},
            \href{mailto:wulff@physics.muni.cz}{wulff@physics.muni.cz}
             \\
		\end{flushleft}

		\section*{Abstract}
		{
			The one-loop correction to eleven-dimensional supergravity involves a cubically divergent term $t_8t_8R^4$, with four Riemann tensors. A similar term (with finite coefficient) has been argued to be present in the M-theory effective action. It is expected to reduce to a similar one-loop term present in the type IIA effective action. This has previously been verified in the NSNS sector at the 4-point level. Here we extend this result to couplings of NSNS and RR fields, which have been computed using the pure spinor formalism. In particular, we check all couplings of RR fields to the dilaton as well as all couplings involving the metric and RR three-form. Correcting some minor mistakes in the literature we find complete agreement. We also give a complete analysis of 4-point terms in eleven dimensions computed previously from superparticle amplitudes and present a very simple form for these.
		}
		
	\end{samepage}
	
	\newpage
	
	\tableofcontents\thispagestyle{fancy}

\section{Introduction} \label{sec:intro}
Eleven-dimensional supergravity \cite{Cremmer:1978km} is UV divergent at one loop. The first divergence occurs in the 4-point scattering of gravitons \cite{Fradkin:1982kf} and corresponds to a term in the effective action of the form\footnote{We will denote fields in eleven dimensions with a hat to distinguish them from their ten-dimensional counterparts. The definition of $t_8$ is given in (\ref{eq:t8}).}
\begin{equation}
\Lambda^3t_8t_8\hat R^4\,,
\end{equation}
with $\Lambda$ the UV cut-off. A similar, quadratically divergent, term arises in type IIA supergravity. This divergence is regulated in string theory and the type IIA string effective action has a similar term at one loop, with \emph{finite} coefficient proportional to $\zeta(2)$. Similarly, it is expected that the M-theory effective action will again contain a similar term, whose  dimensional reduction should reproduce the one-loop type IIA term (these terms are expected to be related to anomaly canceling $\hat C_3\hat R^4$ and $B_2R^4$ terms respectively \cite{Vafa:1995fj,Duff:1995wd}). This term can also be related to a similar term in type IIB at one loop, to similar corrections at tree-level in type IIA and type IIB and even to instanton corrections in type IIB \cite{Green:1997di,Green:1997as,Russo:1997mk}. 

So far we talked only about the metric, but the maximal supersymmetry requires, besides fermions, also other bosonic fields to be present: the three-form in $D=11$ and the dilaton, $B$-field and Ramond-Ramond (RR) fields in $D=10$. It is expected that the complete one-loop correction in $D=11$ will reduce precisely to the one-loop correction to the type IIA string effective action.\footnote{The form of these corrections is not fully known beyond the 4-point level. At higher points there are only partial results, e.g. \cite{Hyakutake:2006aq,Hyakutake:2007sm} and \cite{Richards:2008jg,Richards:2008sa,Liu:2013dna,Liu:2022bfg}. See \cite{Ozkan:2024euj} for a review of higher derivative corrections from the point of view of supersymmetry.} So far this has been checked for the 4-point terms involving only NSNS sector fields ($G,B,\Phi$). Here we extend this result to the couplings of RR and NSNS sector fields, which have been computed using the pure spinor formalism. In particular, we show that all couplings of RR fields to the dilaton coming from the reduction from $D=11$ match with string amplitudes (after minor corrections of the results in the literature).\footnote{The first version of this paper erroneously claimed a disagreement in these terms. This was due to not correctly accounting for certain Ricci terms produced in the string amplitudes computation. There is some overlap with the paper \cite{Liu:2025uqu}, which appeared after the first version of this paper and which is mainly concerned with the type IIB case.} We also show that all couplings involving the metric and RR three-form match with string amplitudes.

The standard reduction ansatz is
\begin{equation}
\hat e^a=e^{-\Phi/3}e^a\,,\qquad\hat e^{10}=e^{2\Phi/3}(dx^{10}+A)\,,
\qquad\hat F\equiv d\hat C=F_4+H\wedge(dx^{10}+A)\,.
\label{eq:reduction}
\end{equation}
Here $e^a$ is the string frame vielbein, $A$ the RR one-form potential ($F_2=dA$) and $F_4$ the RR four-form field strength. This reduction clearly treats the RR one-form and three-form very differently. It also introduces the dilaton and $10d$ metric in an asymmetric fashion. However, in string theory the RR field strengths always enter together in the RR bispinor, similarly the dilaton and metric always enter together since their vertex operators have a similar form. It seems like it would take a miracle for some complicated correction in $D=11$ to give something with the structure we expect from string theory in $D=10$. This suggests that in general one should not expect this to work. However, in the present case supersymmetry is expected to be strong enough to uniquely fix both corrections, forcing them to match, which is indeed what we find.\footnote{Most works in the literature have considered only the metric and $B$-field.
Exceptions include \cite{Frolov:2001jh}, which showed that (for Einstein spaces) the terms linear in the dilaton in the NSNS sector go away after field redefinitions and \cite{Peeters:2003pv}, which computed certain 4-point RR couplings at one loop. Such couplings were also computed in \citep{Bakhtiarizadeh:2013zia}. } 

In this paper we do the following. We perform the reduction of the $\hat R^4$ and $\hat R^2\hat F^2$ terms in $D=11$ and analyze in detail the couplings of the RR fields to the dilaton. We verify that these agree with one-loop 4-point string amplitude calculations \cite{Schwarz:1982jn,Policastro:2006vt} (after fixing minor mistakes in the literature). Note that the one-loop 4-point string amplitude must be the same as the tree-level one (up to the overall coefficient). This follows from the structure of string scattering amplitudes. It also follows from S-duality, in type IIB, and T-duality to type IIA. This is in fact what we check, since we compare to the tree-level pure spinor calculation of \cite{Policastro:2006vt}.

For completeness we also construct the 4-point terms in $D=11$ by uplifting the one-loop NSNS 4-point terms and requiring agreement with 4-point superparticle amplitudes computed by Peeters, Plefka and Stern (PPS) in \cite{Peeters:2005tb}.\footnote{Terms of the form $\hat F^3\hat R$, which are not captured by light-cone superparticle amplitudes, are in principle possible. We will not consider such terms, as their presence does not seem compatible with string amplitudes.}  We find a very simple form for the 4-point $D=11$ Lagrangian, eq. (\ref{eq:Lhats}), and correct a factor of 6 in their result.\\

The outline of the rest of this paper is the following:

\begin{itemize}
    \item[$\diamond$] In section \ref{sec:reduction}, we dimensionally reduce the $\hat R^4$ and $\hat R^2\hat F^2$ terms from eleven to ten dimensions. We analyze in detail the couplings to the dilaton.
    \item[$\diamond$] We then show in section \ref{sec:amp} that the couplings to the dilaton precisely reproduce those coming from one-loop string amplitudes. We also verify this for the $R^2F_2^2$ terms. 
    \item[$\diamond$] Section \ref{sec:uplift} discusses the uplift of the NSNS sector terms to $D=11$ and a simple form for the Lagrangian that results after matching with $D=11$ superparticle amplitudes and from which one directly sees the correct metric and three-form couplings emerging upon dimensional reduction.
    \item[$\diamond$]  We conclude in \pref{sec:conclusion} with some interesting future directions. 
    \item[$\diamond$] Appendices \pref{app:t8s}, \pref{app:PPS}, \pref{app:superparticle} and \pref{app:Weyl} collect several technical details of the calculations in this paper.
\end{itemize}

\section{Reduction to ten dimensions} \label{sec:reduction}
In this section,  we carry out the dimensional reduction of the 4-point terms in the $D=11$ one-loop effective action. Since we are mainly interested in terms involving the dilaton, we will consider only the $\hat R^4$ and $\hat R^2\hat F^2$ terms, ignoring the $\hat F^4$ terms, which give no dilaton terms. Note, however, that the latter are known to give the right NSNS sector terms, $t_8t_8(\nabla H)^4$ \cite{Peeters:2005tb}.

Carrying out the dimensional reduction using the ansatz in (\ref{eq:reduction}) leads to long calculations. The work can be simplified considerably by instead splitting the reduction into three steps:
\begin{itemize}
	\item[1.] Perform a Weyl transformation in $D=11$
	\begin{equation}
	\hat e^a=e^{2\Phi/3}\hat e'^a \; .
    \label{eq:step1}
	\end{equation}
	\item[2.] Perform the reduction using the ansatz
\begin{equation}
\hat e'^a=e'^a\,,\qquad\hat e'^{10}=dx^{10}+A \;.
\label{eq:step2}
\end{equation}
	\item[3.] Perform a Weyl transformation in $D=10$ to go to string frame
	\begin{equation}
	e'^a=e^{-\Phi}e^a \;.
	\label{eq:step3}
\end{equation}
	\end{itemize}
The $B$-field and RR two-form field strength arise, in the second step, via ($H=dB$)
\begin{equation}
    \hat F_{abc10}=H_{abc}\,,\qquad\hat R'_{abc10}=-\tfrac12\nabla_cF_{ab}\,.
\label{eq:red-10-comp}
\end{equation}
The remaining components of $\hat F$ and $\hat R'$ reduce just to $F_4$ and $R'$ in $D=10$.\footnote{Note that, since we are interested only in the 4-point effective action we can drop all higher order terms, in particular, any factor of $e^\Phi$ multiplying other fields.} To avoid cluttering the notation we do not distinguish between $10d$ and $11d$ indices. The range of the index will be clear from the context.

\subsection{Reduction of \texorpdfstring{$\hat R^4$}{R4} terms}
Since we have to perform Weyl transformations of $t_8t_8R^4$ both in ten and eleven dimensions it is convenient to analyze this in the general case. A general Weyl transformation
\begin{equation}
e^a=e^{k\Phi}e'^a
\end{equation}
gives
\begin{equation}
R_{abcd}=e^{-2k\Phi}(R'_{abcd}+D_{abcd}+\ldots)\qquad\mbox{with}\qquad D^{ab}{}_{cd}=4k\delta^{[a}_{[c}\nabla^{\phantom{a}}_{d]}\nabla^{b]}\Phi\,,
\label{eq:RWeyl}
\end{equation}
where the higher order terms, denoted by the ellipsis, will not be relevant for us. In appendix \ref{app:Weyl} we prove that, up to total derivatives, field redefinitions and higher order terms, this leads to
\begin{equation}
t_8t_8R^4
=
t_8t_8R'^4
+96k(2l-k(D-2))\nabla^a\nabla^b\Phi\nabla^c\nabla^d\Phi X_{abcd}
+\ldots
\label{eq:Weyl}
\end{equation}
where 
\begin{equation}
\begin{split}
X_{ab}{}^{cd}=\,&
8R'_{aghb}R'^{cghd}
+8R'_{agh}{}^cR'_b{}^{hgd}
-8\delta_a^dR'_{begh}R'^{cegh}
+\delta_a^c\delta_b^dR'_{efgh}R'^{efgh}
\\
&\quad
+2k(2l-k(D-2))\big[3\nabla_a\nabla_b\Phi\nabla^c\nabla^d\Phi-2\nabla_a\nabla^c\Phi\nabla_b\nabla^d\Phi\big]\,.
\end{split}
\label{eq:X}
\end{equation}
Here $D$ is the dimension and $l$ is given by the linearized equation of motion for the metric
\begin{equation}
R_{ab}=l \, \nabla_a\nabla_b\Phi\,.
\end{equation}

Applying this to the Weyl transformation in $D=11$, equation (\ref{eq:step1}), we have ($k=\frac23$, $l=0$)
\begin{equation}
t_8t_8\hat R^4
=
t_8t_8\hat R'^4
-2^73\nabla^a\nabla^b\Phi\nabla^c\nabla^d\Phi\hat X_{abcd}
+\ldots\,,
\end{equation}
with $\hat X_{abcd}$ given by (\ref{eq:X}), i.e.
\begin{equation}
\begin{split}
\hat X_{abcd}=\,&
8\hat R'_{aghb}\hat R'_c{}^{gh}{}_d
+8\hat R'_{aghc}\hat R'_b{}^{hg}{}_d
-8\eta_{ad}\hat R'_{begh}\hat R'_c{}^{egh}
+\eta_{ac}\eta_{bd}\hat R'_{efgh}\hat R'^{efgh}
\\
&\quad
-24\nabla_a\nabla_b\Phi\nabla_c\nabla_d\Phi
+16\nabla_a\nabla_c\Phi\nabla_b\nabla_d\Phi\,.
\end{split}
\label{eq:Xhat}
\end{equation}

From the definition of $t_8$ in (\ref{eq:t8}) we find that
\begin{equation}
\begin{split}
t_8t_8R^4
=&
12
\big(
16R_{efab}R^{fgbc}R_{ghcd}R^{heda}
+32R_{efab}R^{hebc}R^{fg}{}_{cd}R_{gh}{}^{da}
-16R_{efab}R^{fgab}R_{ghcd}R^{hecd}
\\
&\quad
-8R_{efab}R_{gh}{}^{ab}R^{hecd}R^{fg}{}_{cd}
+R_{abcd}R^{abcd}R_{efgh}R^{efgh}
+2R_{efab}R^{ghab}R^{efcd}R_{ghcd}
\big)\,.
\end{split}
\end{equation}
Using (\ref{eq:red-10-comp}) the reduction of the $D=11$ expression gives
\begin{equation}
t_8t_8R'^4
+L_{R'^2F_2^2}
+L_{F_2^4}
-2^73\nabla^a\nabla^b\Phi\nabla^c\nabla^d\Phi(X_{abcd}+\tilde X_{abcd})\,,
\end{equation}
with
\begin{equation}
\begin{split}
L_{R'^2F_2^2}
&=
12
\big(
-32R'_{efab}R'^{heda}\nabla^fF^{bc}\nabla_hF_{cd}
-32R'_{fgcd}R'^{ghda}\nabla^fF_{ab}\nabla_hF^{bc}
\\
&\quad
-32R'_{hebc}R'^{gh}{}_{da}\nabla^eF^{ab}\nabla_gF^{cd}
-16R'_{ghcd}R'^{hecd}\nabla_aF_{ef}\nabla^aF^{fg}
\\
&\quad
+8R'_{ghcd}R'^{hecd}\nabla_eF_{ab}\nabla^gF^{ab}
+8R'_{efab}R'^{hecd}\nabla^fF^{ab}\nabla_hF_{cd}
\\
&\quad
-8R'_{efab}R'_{gh}{}^{ab}\nabla_cF^{he}\nabla^cF^{fg}
+8R'_{ghab}R'^{fgcd}\nabla_fF^{ab}\nabla^hF_{cd}
\\
&\quad
+4R'_{efab}R'^{efcd}\nabla^gF^{ab}\nabla_gF_{cd}
+2R'_{efab}R'^{efab}\nabla_gF_{cd}\nabla^gF^{cd}
\big)\,,
\end{split}
\label{eq:LR2F2}
\end{equation}
while $X_{abcd}$ is the same as $\hat X_{abcd}$ in (\ref{eq:Xhat}) but with $\hat R'$ replaced by $R'$, while the $F_2$ terms produced in the reduction of this term are given by
\begin{equation}
\tilde X_{abcd}=
4\nabla_aF_{be}\nabla_cF_d{}^e
+4\nabla_aF_{ce}\nabla_dF_b{}^e
-4\eta_{ad}\nabla_eF_{fb}\nabla^eF^f{}_c
-2\eta_{ad}\nabla_bF_{ef}\nabla_cF^{ef}
+\eta_{ac}\eta_{bd}\nabla_gF_{ef}\nabla^gF^{ef}\,.
\end{equation}
Since we will not need them we will not spell out the $F_2^4$ terms.

Finally, performing the $D=10$ Weyl rescaling in (\ref{eq:step3}), using (\ref{eq:Weyl}) with $e$ and $e'$ exchanged and
\begin{equation}
k=-1\,,\qquad l=-\tfrac23(11-2)=-6\,,
\end{equation}
the $X_{abcd}$ term cancels out (using (\ref{eq:D3terms})) and we are left with
\begin{equation}
t_8t_8R^4
+L_{R'^2F_2^2}
+L_{F_2^4}
-2^73\nabla^a\nabla^b\Phi\nabla^c\nabla^d\Phi\tilde X_{abcd}
+\ldots\,.
\end{equation}

Note that the second term contains terms linear and quadratic in the dilaton since, from (\ref{eq:RWeyl}),
\begin{equation}
R'_{abcd}=R_{abcd}-4\delta^{[a}_{[c}\nabla^{\phantom{a}}_{d]}\nabla^{b]}\Phi\,.
\label{eq:Rprime}
\end{equation}
We now analyze these terms in detail.

\subsection{Dilaton couplings}
We have found that upon dimensional reduction (to string frame)
\begin{equation}
t_8t_8\hat R^4\rightarrow
t_8t_8R^4
+L_{R^2F_2^2}
+L_{F_2^4}
+L_{\Phi RF_2^2}
+L_{\Phi^2F_2^2}
+\ldots\,,
\end{equation}
with $L_{R^2F_2^2}$ given by (\ref{eq:LR2F2}) with primes dropped.

Dropping equation of motion terms, which can be removed by field redefinitions, one finds that the terms linear in the dilaton can be simplified drastically to
\begin{equation}
\begin{split}
L_{\Phi RF_2^2}=\,&
2^8 3\nabla_a\nabla^b\Phi
\big(
2R^{aecd}\nabla^fF_{bc}\nabla_eF_{df}
+2R^{ae}{}_{cd}\nabla_bF^{cf}\nabla^dF_{ef}
+R^{ac}{}_{db}\nabla_eF_{cf}\nabla^eF^{fd}
\big)
\\
=&
-2^9 3\Phi R^{ab}{}_{cd}\nabla_a\nabla^eF^{fc}\nabla^d\nabla_fF_{eb}
+\ldots\,,
\end{split}
\label{eq:PhiRF22}
\end{equation}
where, in the last step, we integrated by parts and dropped total derivatives, higher order terms and equation of motion terms. Similarly, the terms quadratic in the dilaton simplify to
\begin{equation}
\begin{split}
L_{\Phi^2F_2^2}=\,&
2^7 3\nabla_a\nabla^b\Phi\,\nabla_c\nabla_d\Phi
\big(
2\nabla^aF^{ce}\nabla^dF_{be}
-2\nabla^aF_{be}\nabla^cF^{de}
+2\nabla_eF^{ac}\nabla^eF_b{}^d
\\
&\qquad\qquad\qquad
+4\eta^{ad}\nabla^eF_{bf}\nabla^fF^{ce}
+3\eta^{ad}\nabla_bF^{ef}\nabla^cF_{ef}
-\eta^{ad}\delta^c_b\nabla_gF_{ef}\nabla^gF^{ef}
\big)
\\
=&
-2^83\Phi\nabla_a\nabla^b\Phi\,\nabla^c\nabla^dF^{ea}\nabla_c\nabla_dF_{eb}
+\ldots\,.
\end{split}
\label{eq:Phi2F22}
\end{equation}
In section \ref{sec:amp} we show that these terms correctly reproduce the corresponding one-loop 4-point type IIA string amplitudes.

\subsection{Reduction of \texorpdfstring{$\hat R^2\hat F^2$}{R2F2} terms}
Again we are primarily interested in the dilaton terms. To compute these it is easiest to start from the expression for the $\hat R^2\hat F^2$ terms given in (\ref{eq:R2dF2}) (see also appendix \ref{app:PPS}). After the Weyl rescaling we get
\begin{equation}
\hat L_{\hat R'^2\hat F^2}
+\hat L_{\Phi\hat R'\hat F^2}
+\hat L_{\Phi^2\hat F^2}
+\ldots\,.
\end{equation}
Here $\hat L_{\hat R'^2\hat F^2}$ is given by (\ref{eq:R2dF2}) but with $\hat R$ replaced by $\hat R'$.

With some work one can show that in general the dilaton terms generated can be simplified to
\begin{equation}
L_{\Phi R' F^2}
=
48k\Phi R'_{abcd}
\big(
-3\nabla^4(F^{abef}F^{cd}{}_{ef})
+8\nabla^2(\nabla^cF^{aefg}\nabla^dF^b{}_{efg})
-4\nabla^a\nabla^cF^{efgh}\nabla^b\nabla^dF_{efgh}
\big)
\label{eq:PhiRF2-gen}
\end{equation}
and
\begin{equation}
\begin{split}
L_{\Phi^2 F^2}
=\,&
16k(2l-k(D-2))\nabla^a\nabla^b\Phi\nabla^c\nabla^d\Phi
\big(
-12\nabla_gF_{acef}\nabla^gF_{bd}{}^{ef}
+8\nabla_aF_{befg}\nabla_cF_d{}^{efg}
\\
&\quad
-8\eta_{ac}\nabla_hF_{befg}\nabla^hF_d{}^{efg}
-2\eta_{ac}\nabla_bF_{efgh}\nabla_dF^{efgh}
+\eta_{ac}\eta_{bd}\nabla_kF_{efgh}\nabla^kF^{efgh}
\big)
\\
&\quad
-288k^2\Phi\nabla_a\nabla^b\Phi\nabla^4\big(F_{befg}F^{aefg}\big)\,.
\end{split}
\label{eq:Phi2F2-gen}
\end{equation}
Setting $D=11$, $k=\frac23$ and $l=0$ we find
\begin{equation}
\hat L_{\Phi\hat R'\hat F^2}
=
32\Phi\hat R'_{abcd}
\big(
-3\nabla^4(\hat F^{abef}\hat F^{cd}{}_{ef})
+8\nabla^2(\nabla^c\hat F^{aefg}\nabla^d\hat F^b{}_{efg})
-4\nabla^a\nabla^c\hat F^{efgh}\nabla^b\nabla^d\hat F_{efgh}
\big)
\label{eq:PhiRF2}
\end{equation}
and
\begin{equation}
\begin{split}
\hat L_{\Phi^2\hat F^2}
=\,&
64\nabla^a\nabla^b\Phi\nabla^c\nabla^d\Phi
\big(
12\nabla_g\hat F_{acef}\nabla^g\hat F_{bd}{}^{ef}
-8\nabla_a\hat F_{befg}\nabla_c\hat F_d{}^{efg}
\\
&\quad
+8\eta_{ac}\nabla_h\hat F_{befg}\nabla^h\hat F_d{}^{efg}
+2\eta_{ac}\nabla_b\hat F_{efgh}\nabla_d\hat F^{efgh}
-\eta_{ac}\eta_{bd}\nabla_k\hat F_{efgh}\nabla^k\hat F^{efgh}
\big)
\\
&\quad
-128\Phi\nabla_a\nabla^b\Phi\nabla^4\big(\hat F_{befg}\hat F^{aefg}\big)\,.
\end{split}
\label{eq:Phi2F2}
\end{equation}

After reduction to $D=10$, eq. (\ref{eq:step2}), we get
\begin{equation}
L_{R'^2H^2}
+L_{\Phi R'H^2}
+L_{\Phi^2H^2}
+L_{R'^2F_4^2}
+L_{\Phi R'F_4^2}
+L_{\Phi^2F_4^2}
+L_{R'HF_2F_4}
+L_{\Phi HF_2F_4}
+L_{H^2F_2^2}
+L_{F_2^2F_4^2}\,,
\end{equation}
where $L_{R'^2F_4^2}$, $L_{\Phi R'F_4^2}$ and $L_{\Phi^2F_4^2}$ are given by (\ref{eq:R2dF2}), (\ref{eq:PhiRF2}) and (\ref{eq:Phi2F2}) upon dropping the hats. The NSNS sector terms are given by
\begin{equation}
\begin{split}
L_{R'^2H^2}
=&
16
\big(
48R'_{abcd}R'_{efg}{}^d\nabla^aH^{cg}{}_h\nabla^bH^{efh}%
-12R'_{abcd}R'_{efg}{}^d\nabla^cH^{ab}{}_h\nabla^gH^{efh}%
\\
&\quad
-12R'_{abcd}R'_{efg}{}^d\nabla_iH^{efc}\nabla^iH^{abg}%
+12R'_{abcd}R'_{efg}{}^d\nabla^cH^{ef}{}_h\nabla^gH^{abh}%
\\
&\quad
+24R'_{abcd}R'^{eafc}\nabla_iH^b{}_{fg}\nabla^iH_e{}^{dg}%
+24R'_{abcd}R'^{eafc}\nabla_iH^{bdg}\nabla^iH_{efg}%
\\
&\quad
+24R'_{abcd}R'^{eafc}\nabla^bH^{dgh}\nabla_eH_{fgh}%
-12R'_{abcd}R'^{efcd}\nabla^aH_{egh}\nabla^bH_f{}^{gh}%
\\
&\quad
+12R'_{abcd}R'^{eacd}\nabla_iH^{bgh}\nabla^iH_{egh}%
+4R'_{abcd}R'_{eacd}\nabla^bH_{fgh}\nabla^eH^{fgh}%
\\
&\quad
+3R'_{abcd}R'^{abcd}\nabla_eH_{fgh}\nabla^fH^{egh}%
\big)\,,
\end{split}
\label{eq:R2H2}
\end{equation}
\begin{equation}
L_{\Phi R'H^2}
=
64\Phi R'_{abcd}
\big(
-3\nabla^4(H^{abe}H^{cd}{}_e)
+12\nabla^2(\nabla^cH^{aef}\nabla^dH^b{}_{ef})
-8\nabla^a\nabla^cH^{efg}\nabla^b\nabla^dH_{efg}
\big)
\end{equation}
and
\begin{equation}
\begin{split}
L_{\Phi^2H^2}
=\,&
2^8\nabla^a\nabla^b\Phi\nabla^c\nabla^d\Phi
\big(
6\nabla_gH_{ace}\nabla^gH_{bd}{}^e
-6\nabla_aH_{bef}\nabla_cH_d{}^{ef}
\\
&\quad
+6\eta_{ac}\nabla_hH_{bef}\nabla^hH_d{}^{ef}
+2\eta_{ac}\nabla_bH_{efg}\nabla_dH^{efg}
-\eta_{ac}\eta_{bd}\nabla_kH_{efg}\nabla^kH^{efg}
\big)
\\
&\quad
-2^73\Phi\nabla_a\nabla^b\Phi\nabla^4\big(H_{bef}H^{aef}\big)\,.
\end{split}
\end{equation}
While the $F_2F_4$ terms are given by
\begin{equation}
\begin{split}
L_{R'HF_2F_4}
=&
16
\big(
24R'_{abcd}\nabla^hH^{cef}\nabla_eF^{gd}\nabla_hF^{ab}{}_{fg}
-12R'^{ab}{}_{cd}\nabla^fH_{abe}\nabla^cF_{gh}\nabla_fF^{degh}
\\
&\quad
-12R'_{abcd}\nabla^eH_{fgh}\nabla^fF^{ec}\nabla^dF^{abgh}
-12R'_{abcd}\nabla^eH^{cgh}\nabla^dF_{ef}\nabla^fF^{ab}{}_{gh}
\\
&\quad
+12R'_{abcd}\nabla^cH_{egh}\nabla^eF^{df}\nabla_fF^{abgh}
+12R'_{abcd}\nabla^aH^{cef}\nabla^dF^{gh}\nabla^bF_{efgh}
\\
&\quad
-24R'_{abcd}\nabla^aH^{egh}\nabla_eF_f{}^c\nabla^bF^{dfgh}
+24R'_{ab}{}^{cd}\nabla^eH^{agh}\nabla_cF^{bf}\nabla_eF_{fdgh}
\\
&\quad
-6R'_{abcd}\nabla^eH_{fgh}\nabla^fF^{ab}\nabla_eF^{cdgh}
-8R'_{abcd}\nabla_eH_{fgh}\nabla^cF^{ea}\nabla^bF^{dfgh}
\\
&\quad
+4R'_{abcd}\nabla^aH_{fgh}\nabla^eF^{cd}\nabla^bF_{efgh}
-4R'_{abcd}\nabla_eH_{fgh}\nabla^aF^{cd}\nabla^eF^{bfgh}
\big)
\end{split}
\label{eq:RHF2F4}
\end{equation}
and
\begin{equation}
L_{\Phi HF_2F_4}
=
64\Phi 
\big[
3\nabla^eF_{ab}\nabla^4(F^{abcd}H_{ecd})
+4\nabla_fF_{ab}\nabla^2(\nabla^aF^{fcde}\nabla^bH_{cde})
\big]\,.
\end{equation}
We do not spell out $L_{H^2F_2^2}$ and $L_{F_2^2F_4^2}$ as we will not use them.

Finally, we perform the Weyl rescaling in $D=10$, equation (\ref{eq:step3}), to go to the string frame. A relatively short calculation shows that the dilaton terms generated in the NSNS sector precisely cancel against the terms above, leaving no dilaton couplings in the NSNS sector, in agreement with string amplitudes. However, for terms involving the RR fields this cancellation does not happen.

Finally, the reduction of the $\hat R^2\hat F^2$ terms to $D=10$ gives (in string frame)
\begin{equation}
L_{R^2H^2}
+L_{R^2F_4^2}
+L_{RHF_2F_4}
+L_{\Phi RF_4^2}
+L_{\Phi^2F_4^2}
+L_{\Phi HF_2F_4}
+L_{H^2F_2^2}
+L_{F_2^2F_4^2}\,,
\end{equation}
with the first three given by (\ref{eq:R2H2}), (\ref{eq:R2dF2}) and (\ref{eq:RHF2F4}) with primes removed. Using  (\ref{eq:PhiRF2-gen}) with $k=-1$ and (\ref{eq:Rprime}) the terms involving the dilaton simplify, upon integrating by parts and dropping equation of motion and higher order terms, to
\begin{equation}
L_{\Phi RF_4^2}
=
16\Phi R_{abcd}
\big(
3\nabla^4(F^{abef}F^{cd}{}_{ef})
-8\nabla^2(\nabla^cF^{aefg}\nabla^dF^b{}_{efg})
+4\nabla^a\nabla^cF^{efgh}\nabla^b\nabla^dF_{efgh}
\big)\,,
\label{eq:PhiRF42}
\end{equation}
\begin{equation}
L_{\Phi^2F_4^2}
=
-32\Phi\nabla_a\nabla^b\Phi\,\nabla^4\big(F^{acde}F_{bcde}\big)
\label{eq:Phi2F42}
\end{equation}
and
\begin{equation}
L_{\Phi HF_2F_4}
=
64\Phi
\big[
3\nabla_aH_{bef}\nabla^2(\nabla^bF_{gh}\nabla^aF^{efgh})
-2\nabla_aH_{cde}\nabla^2(\nabla^fF^{ag}\nabla_gF_{fcde})
\big]\,.
\label{eq:PhiHF2F4}
\end{equation}
In the next section we verify that this precisely reproduces the corresponding 4-point one-loop type IIA string scattering amplitudes.

\section{Comparison to string amplitudes}\label{sec:amp}
Having performed the dimensional reduction in the previous section we will now verify that the reduced action correctly reproduces the one-loop 4-point S-matrix of type IIA string theory. Since the NSNS terms are already known to have the correct structure we focus on the RR terms. Specifically, we will verify that all couplings involving the dilaton, as well as the $R^2F_2^2$ couplings, match with string amplitudes (after correcting a small mistake in the literature).  In the next section we also verify this for the $R^2F_4^2$ and $F_4^4$ couplings.

The type II string spacetime action that reproduces all 4-point \emph{tree-level} amplitudes for external massless states was computed in \citep{Policastro:2006vt}. However, the reduction from $D=11$ should match the one-loop action. But it is known since the early days of string theory that 4-point amplitudes of type II strings, for \textit{all} external massless states, have identical kinematical factor \citep{Schwarz:1982jn,Green:1982sw} (also see \citep{DHoker:1988pdl}), therefore the tree-level and one-loop 4-point amplitudes only differ in the overall factor (which we are ignoring anyway). Therefore we may compare our result to the one of \citep{Policastro:2006vt}. There the action was given in the Einstein frame since it was derived from an amplitude computation. Rewriting it in the string frame, the terms mixing the (bosonic) NSNS and RR sector fields take the form
\begin{equation}
L_{\mathrm{NSNSRR}}=c(A_1+\tfrac12A_2+\tfrac14A_3)
\end{equation}
with
\begin{align}
A_1
=\,&
\mathcal R_{aefb}\mathcal R^{cef}{}_d\tr(\gamma^a\nabla_c\slashed F\gamma^{(b}\nabla^{d)}\slashed F^T)\,,
\label{eq:A1}
\\
A_2
=\,&
\mathcal R_{abeg}\mathcal R_{cdf}{}^g\tr(\gamma^{abc}\nabla^d\slashed F\gamma^{(e}\nabla^{f)}\slashed F^T)
+\mathcal R_{egab}\mathcal R_f{}^g{}_{cd}\tr(\gamma^{abc}\nabla^d\slashed F^T\gamma^{(e}\nabla^{f)}\slashed F)\,,
\\
A_3
=\,&
\mathcal R_{abcd}\mathcal R_{efgh}\tr(\gamma^{abe}\nabla^f\slashed F\gamma^{[cdg}\nabla^{h]}\slashed F^T)\,,
\label{eq:A3}
\end{align}
where $\gamma^a$ are the standard 16-component gamma matrices and we define
\begin{equation}
\mathcal R_{abcd}=R_{abcd}+\nabla_{[a}H_{b]cd}
\end{equation}
and the type IIA RR bispinor
\begin{equation}
\slashed F=\tfrac12F_{ab}\gamma^{ab}+\tfrac{1}{4!}F_{abcd}\gamma^{abcd}\,,\qquad
\slashed F^T=-\tfrac12F_{ab}\gamma^{ab}+\tfrac{1}{4!}F_{abcd}\gamma^{abcd}\,.
\end{equation}
Below we will fix the overall coefficient to be $c=-24$ (we set the coefficient of $t_8t_8R^4$ to 1). Note that the expressions given here differ from those in \cite{Policastro:2006vt} in two respects. First, their normalization of $\slashed F$ is different from ours. Second, and more importantly, their expression for $A_2$ has the same index structure on the $\mathcal R$'s in the two terms, which is not compatible with the parity symmetry of type IIA, so we have corrected this by modifying the second term.

We now check all couplings involving the dilaton. The dilaton does not appear at all in the above expression. However, evaluating the gamma-traces will produce terms involving the (generalized) Ricci tensor, which may be replaced by $R_{ab}=2\nabla_a\nabla_b\Phi$, up to equation of motion terms removable by field redefinitions (the Ricci scalar terms are directly proportional to the equations of motion). Dilaton terms can arise only from contractions of the explicit gamma matrices in (\ref{eq:A1})--(\ref{eq:A3}) and one finds (dropping Ricci scalar terms)
\begin{align}
[A_1]_\Phi
=\,&
-\tfrac12R^{ef}\mathcal R_{cefd}\tr(\nabla^c\slashed F\nabla^d\slashed F^T)\,,
\\
[A_2]_\Phi
=\,&
R_{ag}\mathcal R_{cdf}{}^g\tr(\gamma^{ac}\nabla^d\slashed F\nabla^f\slashed F^T)
+R_{ag}\mathcal R_f{}^g{}_{cd}\tr(\gamma^{ac}\nabla^d\slashed F^T\nabla^f\slashed F)
\nonumber
\\
&\quad
+\tfrac12R_{dg}\mathcal R_{abe}{}^g\tr(\gamma^{ab}\nabla^d\slashed F\nabla^e\slashed F^T)
+\tfrac12R_{dg}\mathcal R_e{}^g{}_{ab}\tr(\gamma^{ab}\nabla^d\slashed F^T\nabla^e\slashed F)\,,
\\
[A_3]_\Phi
=\,&
2R_{ad}\mathcal R_{efgh}\tr(\gamma^{ae}\nabla^f\slashed F\gamma^{dg}\nabla^h\slashed F^T)
+\tfrac12R_{fh}\mathcal R_{abcd}\tr(\gamma^{ab}\nabla^f\slashed F\gamma^{cd}\nabla^h\slashed F^T)
\nonumber
\\
&\quad
+R_{ad}\mathcal R_{efgh}\tr(\gamma^{ae}\nabla^f\slashed F\gamma^{gh}\nabla^d\slashed F^T)
+R_{fh}\mathcal R_{abcd}\tr(\gamma^{ab}\nabla^f\slashed F\gamma^{hc}\nabla^d\slashed F^T)
\nonumber
\\
&\quad
+2R_{ad}R_{fh}\tr(\gamma^a\nabla^f\slashed F\gamma^d\nabla^h\slashed F^T)
-2R_{ad}R_{fh}\tr(\gamma^a\nabla^f\slashed F\gamma^h\nabla^d\slashed F^T)\,.
\end{align}
Evaluating these expressions, dropping equation of motion terms and higher order terms, one finds 
the $F_2F_2$ terms
\begin{align}
[A_1]_{\Phi F_2^2}
=\,&
8\nabla^a\nabla^b\Phi\,R_{acbd}\nabla^cF_{ef}\nabla^dF^{ef}\,,
\\
[A_2]_{\Phi F_2^2}
=\,&
32\nabla_a\nabla^b\Phi
\big(
R_{becd}\nabla^eF^{fa}\nabla_fF^{cd}
+2R_{becd}\nabla_fF^{ac}\nabla^eF^{fd}
\big)\,,
\\
[A_3]_{\Phi F_2^2}
=\,&
32\nabla^a\nabla^b\Phi
\big(
R_{acbd}\nabla^cF_{ef}\nabla^dF^{ef}
-4R^{cdef}\nabla_aF_{be}\nabla_fF_{cd}
+4R^{cdef}\nabla_eF_{ca}\nabla_dF_{fb}
\nonumber\\
&\quad
-R_{cdef}\nabla_aF^{cd}\nabla_bF^{ef}
-2R_{becd}\nabla^eF_{af}\nabla^fF^{cd}
+4R_{be}{}^{cd}\nabla^fF_{ac}\nabla^eF_{df}
\nonumber\\
&\quad
+8R_{be}{}^{cd}\nabla_aF_{fc}\nabla_dF^{fe}
\big)
+64\nabla^a\nabla^b\Phi\,\nabla^c\nabla_d\Phi
\big(
4\nabla_aF_{ce}\nabla_bF^{de}
\nonumber\\
&\quad
-2\nabla_aF_{be}\nabla_cF^{de}
-2\nabla_cF_{ae}\nabla_bF^{de}
+\delta_a^d\nabla_bF_{ef}\nabla_cF^{ef}
\big)\,,
\end{align}
the $F_4F_4$ terms
\begin{align}
[A_1]_{\Phi F_4^2}
=\,&
\tfrac23\nabla^a\nabla^b\Phi\,R_{acbd}\nabla^cF_{efgh}\nabla^dF^{efgh}\,,
\\
[A_2]_{\Phi F_4^2}
=\,&
\tfrac{32}{3}\nabla^a\nabla^b\Phi\,R_{becd}
\big(
\nabla^eF_{afgh}\nabla^cF^{dfgh}
-\nabla^cF_{afgh}\nabla^eF^{dfgh}
+\nabla_aF^{cfgh}\nabla^eF^d{}_{fgh}
\big)\,,
\\
[A_3]_{\Phi F_4^2}
=\,&
\tfrac83\nabla^a\nabla^b\Phi
\big(
R_{acbd}\nabla^cF_{efgh}\nabla^dF^{efgh}
-12R_{cdef}\nabla_aF^{cdgh}\nabla_bF^{ef}{}_{gh}
-24R^{cdef}\nabla_eF_{cagh}\nabla_dF_{bf}{}^{gh}
\nonumber\\
&\quad
-24R^{cdef}\nabla_cF_{dagh}\nabla_eF_{fb}{}^{gh}
-24R^{cdef}\nabla_eF_{fagh}\nabla_bF_{cd}{}^{gh}
-24R^{cdef}\nabla_aF_{begh}\nabla_fF_{cd}{}^{gh}
\nonumber\\
&\quad
+8R_{becd}\nabla^eF_{afgh}\nabla^cF^{dfgh}
+8R_{becd}\nabla^cF_{afgh}\nabla^eF^{dfgh}
+8R_{becd}\nabla_aF^{cfgh}\nabla^eF^d{}_{fgh}
\nonumber\\
&\quad
-16R_{becd}\nabla_aF^{efgh}\nabla^cF^d{}_{fgh}
\big)
+\tfrac{16}{3}\nabla^a\nabla_b\Phi\,\nabla^c\nabla^d\Phi
\big(
8\nabla_cF_{aefg}\nabla_dF^{befg}
\nonumber\\
&\quad
-4\nabla_aF_{cefg}\nabla_dF^{befg}
-4\nabla_aF^{befg}\nabla_cF_{defg}
+\eta_{ad}\nabla_cF_{efgh}\nabla^bF^{efgh}
\big)
\end{align}
and the $F_2F_4$ terms
\begin{align}
[A_1]_{\Phi F_2F_4}
=\,&
0\,,
\\
[A_2]_{\Phi F_2F_4}
=\,&
16\nabla_a\nabla^b\Phi\,(\nabla H)_{becd}
\big(
2\nabla^cF_{gh}\nabla^eF^{adgh}
-2\nabla^eF_{gh}\nabla^cF^{adgh}
-\nabla^aF_{gh}\nabla^eF^{cdgh}
\nonumber\\
&\qquad\qquad
+\nabla^eF_{gh}\nabla^aF^{cdgh}
\big)\,,
\\
[A_3]_{\Phi F_2F_4}
=\,&
64\nabla^a\nabla^b\Phi
\big(
\tfrac13\nabla_fH_{cde}\nabla_aF_{bg}\nabla^fF^{gcde}
-\tfrac23\nabla^fH_{cde}\nabla_aF_{fg}\nabla_bF^{gcde}
\nonumber\\
&\qquad\qquad
+\tfrac13\nabla_fH_{cde}\nabla^fF_{ag}\nabla_bF^{gcde}
+4(\nabla H)^{cdef}\nabla_eF_{ag}\nabla^cF_{bfd}{}^g
\nonumber\\
&\qquad\qquad
-2(\nabla H)^{cdef}\nabla_aF_{cg}\nabla^eF_{bfd}{}^g
-2(\nabla H)^{cdef}\nabla_eF_{cg}\nabla^aF_{bfd}{}^g
\nonumber\\
&\qquad\qquad
+(\nabla H)^{cdef}\nabla_aF_{eg}\nabla_fF_{bcd}{}^g
-(\nabla H)^{cdef}\nabla_eF_{fg}\nabla_aF_{bcd}{}^g
\nonumber\\
&\qquad\qquad
+(\nabla H)_{becd}\nabla^eF^{gh}\nabla^cF_a{}^{dgh}
+(\nabla H)_{becd}\nabla^cF_{gh}\nabla^eF_a{}^{dgh}
\nonumber\\
&\qquad\qquad
+(\nabla H)_{becd}\nabla_aF_{gh}\nabla^cF^{degh}
+(\nabla H)_{becd}\nabla^cF_{gh}\nabla_aF^{degh}
\big)\,.
\end{align}
These expressions look complicated, but they simplify a lot after integration by parts and it is not hard to show that
\begin{equation}
-24(A_1+\tfrac12A_2+\tfrac14A_3)_\Phi
=
L_{\Phi RF_2^2}
+L_{\Phi^2F_2^2}
+L_{\Phi RF_4^2}
+L_{\Phi^2F_4^2}
+L_{\Phi HF_2F_4}
+\ldots\,,
\end{equation}
reproducing precisely the expressions from dimensional reduction in (\ref{eq:PhiRF22}), (\ref{eq:Phi2F22}), (\ref{eq:PhiRF42}), (\ref{eq:Phi2F42}) and (\ref{eq:PhiHF2F4}). This completes the check of all dilaton couplings.

Finally, let us check some couplings without the dilaton. The simplest are the $R^2F_2^2$ terms. Using the expressions (\ref{eq:A1})--(\ref{eq:A3}) one finds, after some algebra, that indeed
\begin{equation}
-24(A_1+\tfrac12A_2+\tfrac14A_3)_{R^2F_2^2}
=
L_{R^2F_2^2}
+12\cdot6!R_{[ab}{}^{ab}R_{cd}{}^{cd}\nabla_eF^{eg}\nabla_{f]}F^f{}_g
+24\cdot6!R_{[ab}{}^{ag}R_{cd}{}^{bc}\nabla_eF^{de}\nabla_{f]}F^f{}_g\,,
\end{equation}
with $L_{R^2F_2^2}$ given by (\ref{eq:LR2F2}) with the primes removed. As the last two terms are total derivatives at this order this demonstrates the matching of these terms as well.

\section{\texorpdfstring{$D=11$}{D=11} Lagrangian via uplift} \label{sec:uplift}
For completeness, we construct here the 4-point terms in eleven dimensions by uplifting the NSNS sector terms in $D=10$. The resulting Lagrangian has undetermined coefficients, which are fixed by comparing to the superparticle scattering amplitudes of PPS \cite{Peeters:2005tb}. In this way we are able to find a much simpler action than the one presented there (and fix a missing factor of 6).

The type II NSNS sector 4-point terms are most conveniently written using the torsionful Riemann tensor, defined as
\begin{equation}
\mathcal R_{abcd}=R_{abcd}+\nabla_{[a}H_{b]cd}+\tfrac12H_{a[c}{}^eH_{d]be}\,.
\end{equation}
Since we will only work at the 4-point level only the first two terms will be relevant for us. In ten dimensions the only contribution to the 4-point Lagrangian in the NSNS sector is\footnote{At this level the correction is the same at tree-level and one loop for type IIA and IIB. The term $\varepsilon_8\varepsilon_8\mathcal R^4$, which is also present, starts contributing at the 5-point level, since the leading piece is a total derivative.}
\begin{equation}
t_8t_8\mathcal R^4\equiv t_{a_1\cdots a_8}t^{b_1\cdots b_8}\mathcal R^{a_1a_2}{}_{b_1b_2}\cdots\mathcal R^{a_7a_8}{}_{b_7b_8}\,,
\end{equation}
where $t_8$ is defined in the standard way as (for $M_i$ four anti-symmetric matrices)
\begin{equation}
t_8M_1M_2M_3M_4
=
8\tr(M_1M_2M_3M_4)
-2\tr(M_1M_2)\tr(M_3M_4)
+\mathrm{cycl}(2,3,4)\,.
\label{eq:t8}
\end{equation}
One can of course expand this expression out and uplift, term by term, to $D=11$ (see for example \cite{Liu:2013dna,Ciafardini:2024ujx}). However, this does not lead to very nice expressions. In order to simplify the uplifting we will take a different approach, where we work with the fields contracted with gamma matrices instead. In particular, we define
\begin{equation}
\slashed{\mathcal R}=\tfrac14R_{abcd}\Gamma^{ab}\otimes\Gamma^{cd}+\tfrac14\nabla_aH_{bcd}\Gamma^{ab}\otimes\Gamma^{cd}\Gamma_{11}\,.
\label{eq:Rslash}
\end{equation}
Note that we have dropped the $H^2$-term, as it plays no role at the 4-point level. Note also the extra $\Gamma_{11}$ in the second term, which is needed for the uplift since $H_{abc}=\hat F_{abc10}$.

We can define a ``spinor version'' of the $t_8$ tensor as
\begin{equation}
\begin{split}
t_{8s}\slashed M_1\slashed M_2\slashed M_3\slashed M_4
=&
-\tfrac34\str(\slashed M_1\cdots\slashed M_4)
+\tfrac{1}{64}\left[\tr(\slashed M_1\slashed M_2)\tr(\slashed M_3\slashed M_4)+\mathrm{cycl}(2,3,4)\right]
\\
&{}\quad
+\tfrac{1}{256}\left[\tr(\slashed M_1\{\slashed M_2,\hat\Gamma^a\})\tr(\{\hat\Gamma_a,\slashed M_3\}\slashed M_4)+\mathrm{cycl}(2,3,4)\right]\,,
\end{split}
\label{eq:t8s}
\end{equation}
where str denotes the symmetrized trace
\begin{equation}
\str(ABCD)=\tfrac16\tr(ABCD+\mathrm{perm}(B,C,D))
\end{equation}
and the last term involves the $D=11$ gamma matrices $\hat\Gamma^a=(\Gamma^a,\Gamma_{11})$. We have defined it so that, for either $\slashed M_i=\frac12M_i^{ab}\Gamma_{ab}$ or $\slashed M_i=\frac12M_i^{ab}\Gamma_{ab}\Gamma_{11}$, we get
\begin{equation}
t_{8s}\slashed M_1\slashed M_2\slashed M_3\slashed M_4=t_8M_1M_2M_3M_4\,.
\end{equation}
Therefore we can write the 4-point Lagrangian for the NSNS fields as
\begin{equation}
t_{8s}t_{8s}\slashed{\mathcal R}^4\,.
\end{equation}
This simplifies the uplift to $D=11$, which we now turn to. It will be enough to consider only the terms involving the metric and $B$-field, which simplifies the calculations.

First we define the following $D=11$ analog of $\slashed{\mathcal R}$ in (\ref{eq:Rslash})
\begin{equation}
\hat{\slashed{\mathcal R}}=\tfrac14\hat R_{abcd}\hat\Gamma^{ab}\otimes\hat\Gamma^{cd}+\nabla_a\hat F_{bcde}\left(\tfrac{1}{12}\hat\Gamma^{ab}\otimes\hat\Gamma^{cde}+\tfrac{a}{48}\hat\Gamma^a\otimes\hat\Gamma^{bcde}\right)\,,
\end{equation}
where $a$ is a constant to be fixed below. This does not quite reduce to $\slashed{\mathcal R}$, instead we get
\begin{equation}
\hat{\slashed{\mathcal R}}
\rightarrow
\slashed{\mathcal R}
+\slashed{\mathcal R}'
\qquad\mbox{with}\qquad
\slashed{\mathcal R}'=
\tfrac{1}{12}\nabla_aH_{bcd}\left(-\Gamma^a\Gamma_{11}\otimes\Gamma^{bcd}+a\Gamma^a\otimes\Gamma^{bcd}\Gamma_{11}\right)\,.
\end{equation}
Therefore we get (note that only even powers of $\slashed{\mathcal R}'$ can enter, since the trace of an odd number of $\Gamma$'s vanishes)
\begin{equation}
t_{8s}t_{8s}\hat{\slashed{\mathcal R}}^4
\rightarrow
t_{8s}t_{8s}\slashed{\mathcal R}^4
+6t_{8s}t_{8s}\slashed{\mathcal R}^2\slashed{\mathcal R}'^2
+t_{8s}t_{8s}\slashed{\mathcal R}'^4\,.
\label{eq:hatRred}
\end{equation}
We only want the first of these terms and we need to cancel the second and third term somehow. Using the identities in appendix \ref{app:t8s} the second term becomes
\begin{equation}
\begin{split}
6t_{8s}t_{8s}\slashed{\mathcal R}^2\slashed{\mathcal R}'^2
&=
-\tfrac{3}{32}(1+a^2)R_{abcd}R_{efgh}t_{8s}(\Gamma^{ab}\Gamma^{ef}\Gamma^i\Gamma^j)t_{8s}(\Gamma^{cd}\Gamma^{gh}\nabla_i\slashed H\nabla_j\slashed H)
\\
&\quad
-\tfrac{3}{32}(1+a^2)(\nabla H)_{abcd}(\nabla H)_{efgh}t_{8s}(\Gamma^{ab}\Gamma^{ef}\Gamma^i\Gamma^j)t_{8s}(\Gamma^{cd}\Gamma_{11}\Gamma^{gh}\Gamma_{11}\nabla_i\slashed H\nabla_j\slashed H)
\\
&\quad
+192a^2R^a{}_{bcd}R^{bgef}\nabla_{(a}H^{hcd}\nabla_{g)}H_{hef}
+48a^2R_{abcd}R^{abef}\nabla^gH^{hcd}\nabla_gH_{hef}\,,
\end{split}
\end{equation}
where $\slashed H\equiv\frac16H_{abc}\Gamma^{abc}$. We see that the simplest option is to take
\begin{equation}
a=i
\end{equation}
so that the first two terms vanish (the Lagrangian is still real since only even powers of $a$ appear). We still need to cancel the third and fourth term. This can be easily done by adding to the $D=11$ Lagrangian the terms
\begin{equation}
\begin{split}
\hat L_1
=&
96\hat R^a{}_{bcd}\hat R^{bgef}\nabla_{(a}\hat F^{cdij}\nabla_{g)}\hat F_{efij}
+24\hat R_{abcd}\hat R^{abef}\nabla^g\hat F^{cdij}\nabla_g\hat F_{efij}
\\
=&
-\tfrac98t_{8s}(\hat\Gamma^{ab}\hat\Gamma^{cd}\hat\Gamma^e\hat\Gamma^f)\hat R_{abgh}\hat R_{cdij}\nabla_e\hat F^{ghkl}\nabla_f\hat F^{ij}{}_{kl}\,,
\end{split}
\end{equation}
which, upon reduction, cancels the unwanted terms.

Finally, we have the last term in (\ref{eq:hatRred}). Using the identities in appendix \ref{app:t8s} this term becomes
\begin{equation}
\begin{split}
t_{8s}t_{8s}\slashed{\mathcal R}'^4
=&
-3\tr(\nabla_a\slashed H\nabla^a\slashed H\nabla_b\slashed H\nabla^b\slashed H)
+\tfrac32\tr(\nabla_a\slashed H\nabla_b\slashed H\nabla^a\slashed H\nabla^b\slashed H)
\\
&\quad
+\tfrac{3}{32}\tr(\nabla_a\slashed H\nabla^a\slashed H)\tr(\nabla_b\slashed H\nabla^b\slashed H)\,.
\end{split}
\end{equation}
These terms can again be easily canceled by adding the following terms in $D=11$
\begin{equation}
\hat L_2
=
3\tr(\nabla_a\hat{\slashed F}\nabla^a\hat{\slashed F}\nabla_b\hat{\slashed F}\nabla^b\hat{\slashed F})
-\tfrac32\tr(\nabla_a\hat{\slashed F}\nabla_b\hat{\slashed F}\nabla^a\hat{\slashed F}\nabla^b\hat{\slashed F})
-\tfrac{3}{32}\tr(\nabla_a\hat{\slashed F}\nabla^a\hat{\slashed F})\tr(\nabla_b\hat{\slashed F}\nabla^b\hat{\slashed F})\,,
\label{eq:L2}
\end{equation}
with $\hat{\slashed F}\equiv\frac{1}{4!}\hat F_{abcd}\hat\Gamma^{abcd}$.

In summary, we have found that the 4-point terms in the $D=11$ Lagrangian should take the form,
\begin{equation}
\begin{split}
\hat L
=\,&
t_{8s}t_{8s}\hat{\slashed{\mathcal R}}^4
-\tfrac98t_{8s}(\hat\Gamma^{ab}\hat\Gamma^{cd}\hat\Gamma^e\hat\Gamma^f)\hat R_{abgh}\hat R_{cdij}\nabla_e\hat F^{ghkl}\nabla_f\hat F^{ij}{}_{kl}
+3\tr(\nabla_a\hat{\slashed F}\nabla^a\hat{\slashed F}\nabla_b\hat{\slashed F}\nabla^b\hat{\slashed F})
\\
&{}
-\tfrac32\tr(\nabla_a\hat{\slashed F}\nabla_b\hat{\slashed F}\nabla^a\hat{\slashed F}\nabla^b\hat{\slashed F})
-\tfrac{3}{32}\tr(\nabla_a\hat{\slashed F}\nabla^a\hat{\slashed F})\tr(\nabla_b\hat{\slashed F}\nabla^b\hat{\slashed F})
+\hat L'\,,
\end{split}
\label{eq:Lhat}
\end{equation}
to give the correct 4-point terms involving the metric and $B$-field in $D=10$. Note that we have to allow also for terms that do not contribute to pure metric and $B$-field couplings. We denoted these by $\hat L'$. One can show that the most general form for these terms is
\begin{equation}
\hat L'
=
b\hat R_{abcd}\hat R_{efgh}\nabla^k\hat F^{abef}\nabla_k\hat F^{cdgh}
+\sum_i c_iY_i\,,
\label{eq:Lhatp}
\end{equation}
where the independent combinations of $\hat F^4$ terms whose reduction gives no $H^4$ terms are, in the basis of PPS in appendix \ref{app:PPS}, the 10 combinations
\begin{equation}
\begin{split}
Y_1=&C_2+12C_7+4C_9-24C_{10}+6C_{14}+48C_{17}-6C_{18}+24C_{19}-6C_{21}+6C_{23}\,,\\
Y_2=&3C_4-3C_5+C_6+9C_7-6C_8+12C_{17}-3C_{18}\,,\\
Y_3=&3C_{10}+C_{11}-3C_{13}-3C_{15}-6C_{17}+6C_{20}+3C_{21}\,,\\
Y_4=&C_4+3C_7+C_9-6C_{15}+12C_{19}\,,\\
Y_5=&C_1+4C_2-16C_3-16C_6-144C_7+144C_8\,,\\
Y_6=&C_2-6C_4-2C_9+6C_{11}+36C_{12}-18C_{13}+18C_{15}\,,\\
Y_7=&C_2-12C_4+12C_5+4C_9\,,\\
Y_8=&C_9-9C_{15}-C_{22}+9C_{23}\,,\\
Y_9=&C_{21}+C_{24}\,,\\
Y_{10}=&C_{16}-36C_{20}-9C_{21}-C_{22}\,.
\end{split}
\label{eq:Ys}
\end{equation}

In appendix \ref{app:superparticle} we fix the 11 coefficients $b$, $c_i$ by requiring agreement with $D=11$ superparticle amplitudes computed by PPS in \cite{Peeters:2005tb}. This results in the non-zero coefficients
\begin{equation}
b=-48\,,\quad c_2=12\,,\quad c_3=24\,,\quad c_5=\tfrac13\,,\quad c_7=-\tfrac{10}{3}\,,\quad c_8=\tfrac{68}{3}\,,\quad c_9=12\,.
\end{equation}

As mentioned earlier, there is a small disagreement with their result. The two match (up to an overall factor) only if we multiply their $\hat R^2\hat F^2$ terms by 6 and rescale their $\hat F\rightarrow\frac29\hat F$. The missing factor of 6 appears to be due to their calculation not taking into account the symmetry factors ($\frac{1}{4!}$ for $\hat R^4$ and $\hat F^4$ and $\frac14$ for $\hat R^2\hat F^2$).

The resulting Lagrangian, (\ref{eq:Lhat}), does not have a particularly simple form. However, writing it in terms of structures similar to those that arise in $D=10$, one finds (\ref{eq:R2F2s}) and (\ref{eq:F4s}) and the following simple form for the quartic terms in the $D=11$ Lagrangian
\begin{equation}
\begin{split}
\hat L
=\,&
t_8t_8\hat R^4
-\tfrac32\hat R_{abcd}\hat R_{efgh}\tr(\hat\Gamma^c\hat\Gamma^{ef}\nabla^d\hat{\slashed F}\hat\Gamma^a\hat\Gamma^{gh}\nabla^b\hat{\slashed F})
-\tfrac32\hat R_{abcd}\hat R_{efgh}\tr(\hat\Gamma^e\hat\Gamma^{cd}\nabla^f\hat{\slashed F}\hat\Gamma^a\hat\Gamma^{gh}\nabla^b\hat{\slashed F})
\\
&\quad{}
-\tfrac{3}{32}
\tr(\nabla_a\hat{\slashed F}\hat\Gamma_c\nabla_b\hat{\slashed F}\hat\Gamma_d\nabla^a\hat{\slashed F}\hat\Gamma^c\nabla^b\hat{\slashed F}\hat\Gamma^d)
-\tfrac{3}{16}\tr(\nabla_a\hat{\slashed F}\hat\Gamma_c\nabla_b\hat{\slashed F}\nabla^a\hat{\slashed F}\hat\Gamma^c\nabla^b\hat{\slashed F})
\\
&\quad{}
+\tfrac{9}{32}\tr(\nabla_a\hat{\slashed F}\nabla_b\hat{\slashed F}\nabla^a\hat{\slashed F}\nabla^b\hat{\slashed F})\,.
\end{split}
\label{eq:Lhats}
\end{equation}
Besides being much simpler than any form that has appeared previously in the literature, this form of the Lagrangian has the advantage that it is trivial to see that it reproduces the $R^2F_4^2$ and $F_4^4$ terms found in \cite{Policastro:2006vt} upon reduction to type IIA. In fact, we get a simpler expression, involving only two $R^2F_4^2$ terms and one $F_4^4$ term, compared to three $R^2F_4^2$ terms and five $F_4^4$ terms in \cite{Policastro:2006vt}. It is easy to see that our expression is equivalent to theirs.

\section{Future directions} \label{sec:conclusion}
We have shown that the 4-point couplings between the NSNS and RR sector fields that arise from the reduction of the one-loop quartic terms in $D=11$ match with the string amplitude calculations of \cite{Policastro:2006vt} (corrected as in section \ref{sec:amp}). This is consistent with the recent paper \cite{Liu:2025uqu}, which checked the $F_2^2$ terms in a different way. While we have not checked all couplings involving the $B$-field and/or $F_2$ coming from the reduction from $D=11$, given the checks we have performed here, these are essentially guaranteed to match the expressions given in \cite{Policastro:2006vt}, which follow from the double copy of the open string amplitude. 

Interesting future directions include analyzing the supersymmetry of these terms as well as the extension to 5 points, where the tree-level and one-loop amplitudes start to differ. 
We hope to return to these questions in the future.

\section*{Acknowledgement}
We thank G. Bossard, M. Garousi, J. Liu, R. Minasian, R. Savelli, A. Schachner, A. Tseytlin and P. Vanhove for helpful discussions. This work is supported by the Czech Science Foundation GAČR grant ``Dualities and higher derivatives" (GA23-06498S).

\appendix

\section{\texorpdfstring{$t_{8s}$}{t8s} identities}\label{app:t8s}
Here we collect some useful identities satisfied by $t_{8s}$ defined in (\ref{eq:t8s}). Writing
\begin{equation}
\slashed M^{(n)}=\tfrac{1}{n!}M_{a_1\cdots a_n}\Gamma^{a_1\cdots a_n}\,,\qquad\slashed M^{(n')}=\tfrac{1}{n!}M_{a_1\cdots a_n}\Gamma^{a_1\cdots a_n}\Gamma_{11}
\end{equation}
we have
\begin{equation}
\begin{aligned}
t_{8s}(\slashed M_1^{(1')}\slashed M_2^{(1')}\slashed M_3^{(1')}\slashed M_4^{(1')})&=t_{8s}(\slashed M_1^{(1)}\slashed M_2^{(1)}\slashed M_3^{(1)}\slashed M_4^{(1)})\,,\\
t_{8s}(\slashed M_1^{(1)}\slashed M_2^{(1)}\slashed M_3^{(1')}\slashed M_4^{(1')})&=-t_{8s}(\slashed M_1^{(1)}\slashed M_2^{(1)}\slashed M_3^{(1)}\slashed M_4^{(1)})+16M_1^aM_{2a}M_3^bM_{4b}\,,\\
t_{8s}(\slashed M_1^{(2)}\slashed M_2^{(2)}\slashed M_3^{(1')}\slashed M_4^{(1')})&=-t_{8s}(\slashed M_1^{(2)}\slashed M_2^{(2)}\slashed M_3^{(1)}\slashed M_4^{(1)})\,,\\
t_{8s}(\slashed M_1^{(2)}\slashed M_2^{(2)}\slashed M_3^{(3')}\slashed M_4^{(3')})&=-t_{8s}(\slashed M_1^{(2)}\slashed M_2^{(2)}\slashed M_3^{(3)}\slashed M_4^{(3)})+8M_1^{ab}M_{2cd}M_{3abe}M_4^{cde}\,,\\
t_{8s}(\slashed M_1^{(2')}\slashed M_2^{(2')}\slashed M_3^{(3')}\slashed M_4^{(3')})&=
-t_{8s}(\slashed M_1^{(2')}\slashed M_2^{(2')}\slashed M_3^{(3)}\slashed M_4^{(3)})\,,\\
t_{8s}(\slashed M_1^{(3')}\slashed M_2^{(3')}\slashed M_3^{(3')}\slashed M_4^{(3')})&=t_{8s}(\slashed M_1^{(3)}\slashed M_2^{(3)}\slashed M_3^{(3)}\slashed M_4^{(3)})\,,\\
t_{8s}(\slashed M_1^{(3)}\slashed M_2^{(3)}\slashed M_3^{(3')}\slashed M_4^{(3')})&=
-t_{8s}(\slashed M_1^{(3)}\slashed M_2^{(3)}\slashed M_3^{(3)}\slashed M_4^{(3)})
-\tfrac12\tr(\slashed M_1^{(3)}\slashed M_4^{(3)}\slashed M_2^{(3)}\slashed M_3^{(3)})
\\
&\quad
+\tfrac{1}{64}\tr(\slashed M_1^{(3)}\slashed M_3^{(3)})\tr(\slashed M_4^{(3)}\slashed M_2^{(3)})
\\
&\quad
+\tfrac{1}{64}\tr(\slashed M_1^{(3)}\slashed M_4^{(3)})\tr(\slashed M_2^{(3)}\slashed M_3^{(3)})\,.
\end{aligned}
\end{equation}
We also note that
\begin{equation}
t_{8s}(\Gamma^a\Gamma^b\Gamma^c\Gamma^d)=24\eta^{a(b}\eta^{cd)}\,.
\end{equation}

\section{Basis of terms}\label{app:PPS}
For ease of comparison, we use the same basis of terms introduced by PPS \cite{Peeters:2005tb}. The $\hat R^2\hat F^2$ terms are
\begin{equation}
\begin{aligned}
B_{1}  &=\hat R_{abcd}\hat R_{efgh}\nabla^{e}\hat F^{agh}{}_{i}\nabla^{c}\hat F^{bdfi} \hspace{2cm} &
B_{13} &=\hat R_{abcd}\hat R_{e}{}^{a}{}_{f}{}^{c}\nabla_{i}\hat F^{bf}{}_{gh}\nabla^{i}\hat F^{degh}\\
B_{2}  &=\hat R_{abcd}\hat R_{efgh}\nabla^{e}\hat F^{acg}{}_{i}\nabla^{h}\hat F^{bdfi}\hspace{2cm} &
B_{14} &=\hat R_{abcd}\hat R_{e}{}^{a}{}_{f}{}^{c}\nabla_{i}\hat F^{bd}{}_{gh}\nabla^{i}\hat F^{efgh}\\
B_{3}  &=\hat R_{abcd}\hat R_{efgh}\nabla^{e}\hat F^{acg}{}_{i}\nabla^{f}\hat F^{bdhi}\hspace{2cm} &
B_{15} &=\hat R_{abcd}\hat R_{e}{}^{a}{}_{f}{}^{c}\nabla^{b}\hat F^{f}{}_{ghi}\nabla^{e}\hat F^{dghi}\\
B_{4}  &=\hat R_{abcd}\hat R_{efgh}\nabla_{i}\hat F^{cdgh}\nabla^{f}\hat F^{iabe}\hspace{2cm} &
B_{16} &=\hat R_{abcd}\hat R_{e}{}^{a}{}_{f}{}^{c}\nabla^{b}\hat F^{d}{}_{ghi}\nabla^{e}\hat F^{fghi}\\
B_{5}  &=\hat R_{abcd}\hat R_{efg}{}^{d}\nabla^{a}\hat F^{bc}{}_{hi}\nabla^{e}\hat F^{fghi}\hspace{2cm} &
B_{17} &=\hat R_{abcd}\hat R_{e}{}^{a}{}_{f}{}^{c}\nabla^{b}F^{e}{}_{ghi}\nabla^{d}\hat F^{fghi}\\
B_{6}  &=\hat R_{abcd}\hat R_{efg}{}^{d}\nabla^{a}\hat F^{be}{}_{hi}\nabla^{c}\hat F^{fghi}\hspace{2cm} &
B_{18} &=\hat R_{abcd}\hat R_{e}{}^{a}{}_{f}{}^{c}\nabla_{i}\hat F^{ef}{}_{gh}\nabla^{d}\hat F^{bghi}\\
B_{7}  &=\hat R_{abcd}\hat R_{efg}{}^{d}\nabla^{a}\hat F^{be}{}_{hi}\nabla^{g}\hat F^{cfhi}\hspace{2cm} &
B_{19} &=\hat R_{abcd}\hat R_{ef}{}^{cd}\nabla_{i}\hat F^{ae}{}_{gh}\nabla^{i}\hat F^{bfgh}\\
B_{8}  &=\hat R_{abcd}\hat R_{efg}{}^{d}\nabla^{a}F^{ce}{}_{hi}\nabla^{b}\hat F^{fghi}\hspace{2cm} &
B_{20} &=\hat R_{abcd}\hat R_{ef}{}^{cd}\nabla^{a}\hat F^{e}{}_{ghi}\nabla^{b}\hat F^{fghi}\\
B_{9}  &=\hat  R_{abcd}\hat R_{efg}{}^{d}\nabla^{a}\hat F^{ce}{}_{hi}\nabla^{f}\hat F^{bghi}\hspace{2cm} &
B_{21} &=\hat R_{abcd}\hat R_{ef}{}^{cd}\nabla^{a}\hat F^{e}{}_{ghi}\nabla^{f}\hat F^{bghi}\\
B_{10} &=\hat R_{abcd}\hat R_{efg}{}^{d}\nabla_{i}\hat F^{cegh}\nabla^{i}\hat F^{abfh}\hspace{2cm} &
B_{22} &=\hat R_{abcd}\hat R_{e}{}^{acd}\nabla^{b}\hat F_{fghi}\nabla^{e}\hat F^{fghi}\\
B_{11} &=\hat R_{abcd}\hat R_{efg}{}^{d}\nabla_{h}\hat F^{abf}{}_{i}\nabla^{i}\hat F^{cegh}\hspace{2cm} &
B_{23} &=\hat R_{abcd}\hat R_{e}{}^{acd}\nabla_{i}\hat F^{b}{}_{fgh}\nabla^{i}\hat F^{efgh}\\
B_{12} &=\hat R_{abcd}\hat R_{efg}{}^{d}\nabla^{c}\hat F^{ef}{}_{hi}\nabla^{g}\hat F^{bahi}\hspace{2cm} &
B_{24} &=\hat R_{abcd}\hat R^{abcd}\nabla_{e}\hat F_{fghi}\nabla^{f}\hat F^{eghi}
\end{aligned}
\end{equation}
The following 6 combinations vanish modulo total derivatives, equations of motion and higher order terms (i.e. they do not contribute to the 4-point amplitude)
\begin{equation}
\begin{split}
Z_1=&48B_1+48B_2-48B_3+36B_4+96B_6+48B_7-48B_8+96B_{10}\\
&\quad+12B_{12}+24B_{13}-12B_{14}+8B_{15}+8B_{16}-16B_{17}+6B_{19}+2B_{22}+B_{24}\,,\\
Z_2=&-48B_1-48B_2-24B_4-24B_5+48B_6-48B_8-24B_9-72B_{10}\\
&\quad-24B_{13}+24B_{14}-B_{22}+4B_{23}\,,\\
Z_3=&12B_1+12B_2-24B_3+9B_4+48B_6+24B_7-24B_8+24B_{10}+6B_{12}+6B_{13}\\
&\quad+4B_{15}-4B_{17}+3B_{19}+2B_{21}\,,\\
Z_4=&12B_1+12B_2-12B_3+9B_4+24B_6+12B_7-12B_8+24B_{10}+3B_{12}\\
&\quad+6B_{13}+4B_{15}-4B_{17}+2B_{20}\,,\\
Z_5=&4B_3-8B_6-4B_7+4B_8-B_{12}-2B_{14}+4B_{18}\,,\\
Z_6=&B_4+2B_{11}\,.
\end{split}
\label{eq:Brel}
\end{equation}

The $\hat F^4$ terms are
\begin{equation}
\begin{aligned}
C_{1}  &=\nabla_a\hat F_{bcde}\nabla^a\hat F^{bcde}      \nabla_{f}\hat F_{ghij}      \nabla^{f}\hat F^{ghij}\hspace{1.5cm} &
C_{13} &=\nabla_a\hat F_{bcde}\nabla^a\hat F^{b}{}_{fgh} \nabla^{c}\hat F^{fg}{}_{ij} \nabla^{d}\hat F^{ehij}\\
C_{2}  &=\nabla_a\hat F_{bcde}\nabla^a\hat F^{bcd}{}_{f} \nabla^{e}\hat F_{ghij}      \nabla^{f}\hat F^{ghij}\hspace{1.5cm} &
C_{14} &=\nabla_a\hat F_{bcde}\nabla^a\hat F^{b}{}_{fgh} \nabla_{i}\hat F^{cdf}{}_{j} \nabla^{i}\hat F^{eghj}\\
C_{3}  &=\nabla_a\hat F_{bcde}\nabla^a\hat F^{bcd}{}_{f} \nabla_{g}\hat F^{e}{}_{hij} \nabla^{g}\hat F^{fhij}\hspace{1.5cm} &
C_{15} &=\nabla_a\hat F_{bcde}\nabla^a\hat F_{fghi}      \nabla^{b}\hat F^{cdf}{}_{j} \nabla^{g}\hat F^{ehij}\\
C_{4}  &=\nabla_a\hat F_{bcde}\nabla^a\hat F^{bc}{}_{fg} \nabla^{d}\hat F^{e}{}_{hij} \nabla^{f}\hat F^{ghij}\hspace{1.5cm} &
C_{16} &=\nabla_a\hat F_{bcde}\nabla^a\hat F_{fghi}      \nabla^{b}\hat F^{fgh}{}_{j} \nabla^{i}\hat F^{cdej}\\
C_{5}  &=\nabla_a\hat F_{bcde}\nabla^a\hat F^{b}{}_{fgh} \nabla_{i}\hat F^{cde}{}_{j} \nabla^{f}\hat F^{ghij}\hspace{1.5cm} &
C_{17} &=\nabla_a\hat F_{bcde}\nabla^b\hat F^{ac}{}_{fg} \nabla_{h}\hat F^{df}{}_{ij} \nabla^{i}\hat F^{eghj}\\
C_{6}  &=\nabla_a\hat F_{bcde}\nabla^a\hat F^{b}{}_{fgh} \nabla_{i}\hat F^{cde}{}_{j} \nabla^{j}\hat F^{fghi}\hspace{1.5cm} &
C_{18} &=\nabla_a\hat F_{bcde}\nabla^b\hat F^{a}{}_{fgh} \nabla_{i}\hat F^{cdf}{}_{j} \nabla^{j}\hat F^{eghi}\\
C_{7}  &=\nabla_a\hat F_{bcde}\nabla^b\hat F^{ac}{}_{fg} \nabla^{d}\hat F^{e}{}_{hij} \nabla^{h}\hat F^{fgij}\hspace{1.5cm} &
C_{19} &=\nabla_a\hat F_{bcde}\nabla^b\hat F^{c}{}_{fgh} \nabla^{f}\hat F^{dg}{}_{ij} \nabla^{i}\hat F^{aehj}\\
C_{8}  &=\nabla_a\hat F_{bcde}\nabla^b\hat F^{ac}{}_{fg} \nabla_{h}\hat F^{de}{}_{ij} \nabla^{i}\hat F^{fghj}\hspace{1.5cm} &
C_{20} &=\nabla_a\hat F_{bcde}\nabla^b\hat F^{c}{}_{fgh} \nabla^{f}\hat F^{ad}{}_{ij} \nabla^{i}\hat F^{eghj}\\
C_{9}  &=\nabla_a\hat F_{bcde}\nabla^b\hat F_{fghi}      \nabla^{f}\hat F^{ghi}{}_{j} \nabla^{j}\hat F^{acde}\hspace{1.5cm} &
C_{21} &=\nabla_a\hat F_{bcde}\nabla^b\hat F^{c}{}_{fgh} \nabla^{d}\hat F^{fg}{}_{ij} \nabla^{h}\hat F^{aeij}\\
C_{10} &=\nabla_a\hat F_{bcde}\nabla^a\hat F^{bc}{}_{fg} \nabla_{h}\hat F^{df}{}_{ij} \nabla^{i}\hat F^{eghj}\hspace{1.5cm} &
C_{22} &=\nabla_a\hat F_{bcde}\nabla^b\hat F_{fghi}      \nabla^{f}\hat F^{cde}{}_{j} \nabla^{j}\hat F^{aghi}\\
C_{11} &=\nabla_a\hat F_{bcde}\nabla^a\hat F^{bc}{}_{fg} \nabla^{d}\hat F^{f}{}_{hij} \nabla^{g}\hat F^{ehij}\hspace{1.5cm} &
C_{23} &=\nabla_a\hat F_{bcde}\nabla^b\hat F_{fghi}      \nabla^{f}\hat F^{cdg}{}_{j} \nabla^{j}\hat F^{aehi}\\
C_{12} &=\nabla_a\hat F_{bcde}\nabla^a\hat F^{b}{}_{fgh} \nabla^{c}\hat F^{df}{}_{ij} \nabla^{i}\hat F^{eghj}\hspace{1.5cm} &
C_{24} &=\nabla_a\hat F_{bcde}\nabla^b\hat F_{fghi}      \nabla^{f}\hat F^{acd}{}_{j} \nabla^{g}\hat F^{ehij}
\end{aligned}
\end{equation}
The following 9 combinations vanish modulo total derivatives, equations of motion and higher order terms (i.e. they do not contribute to the 4-point amplitude)
\begin{equation}
\begin{split}
\tilde Z_1=&-C_3+12C_4-6C_5+72C_7-9C_8-C_9+54C_{10}-6C_{11}-144C_{12}+18C_{14}-27C_{18}+18C_{21}\,,\\
\tilde Z_2=&C_3-6C_5-18C_7+9C_8+C_9+6C_{11}+9C_{18}+18C_{23}\,,\\
\tilde Z_3=&C_1+96C_4-96C_5+32C_6+288C_7+64C_9+32C_{22}\,,\\
\tilde Z_4=&-C_{10}+2C_{12}+2C_{20}\,,\\
\tilde Z_5=&C_7+C_{10}+4C_{19}\,,\\
\tilde Z_6=&-C_7-C_{10}+2C_{17}\,,\\
\tilde Z_7=&C_1-8C_2+32C_6+32C_9+32C_{16}\,,\\
\tilde Z_8=&-C_2-12C_4+12C_5-4C_9-12C_{11}+36C_{15}\,,\\
\tilde Z_9=&C_{10}-2C_{12}+C_{13}\,.
\end{split}
\label{eq:Crel}
\end{equation}

\section{Comparison to \texorpdfstring{$D=11$}{D=11} superparticle amplitudes (PPS)}\label{app:superparticle}
Here we fix the terms in the $D=11$ Lagrangian not fixed by uplifting the metric an $B$-field terms from $D=10$, i.e. $\hat L'$ in (\ref{eq:Lhat}), by comparing to 4-point superparticle amplitudes computed by PPS in \cite{Peeters:2005tb}.

The first term in the Lagrangian in (\ref{eq:Lhat}) is
\begin{equation}
t_{8s}t_{8s}\hat{\slashed{\mathcal R}}^4
=
t_{8s}t_{8s}\hat{\slashed R}^4
+6t_{8s}t_{8s}\hat{\slashed R}^2\hat{\slashed{\mathcal R}}'^2
+t_{8s}t_{8s}\hat{\slashed{\mathcal R}}'^4\,,
\end{equation}
with
\begin{equation}
\hat{\slashed{\mathcal R}}'=\nabla_a\hat F_{bcde}\left(\tfrac{1}{12}\hat\Gamma^{ab}\otimes\hat\Gamma^{cde}+\tfrac{i}{48}\hat\Gamma^a\otimes\hat\Gamma^{bcde}\right)\,.
\end{equation}
The first term is
\begin{equation}
t_{8s}t_{8s}\hat{\slashed R}^4
=
t_8t_8\hat R^4\,,
\end{equation}
which matches the graviton 4-point function. This, of course, matches with PPS (they have an overall coefficient of $2^{20}$). Next we turn to the $\hat R^2\hat F^2$ terms.

\subsection{\texorpdfstring{$\hat R^2\hat F^2$}{R2F2} terms}
We have
\begin{equation}
\begin{split}
6t_{8s}t_{8s}\hat{\slashed R}^2\hat{\slashed{\mathcal R}}'^2
=&
\tfrac{1}{24}t_{a_1\cdots a_8}\hat R^{a_1a_2}{}_{ab}\hat R^{a_3a_4}{}_{cd}\nabla^{a_5}\hat F^{a_6}{}_{efg}\nabla^{a_7}\hat F^{a_8}{}_{hij}
t_{8s}(\hat\Gamma^{ab}\hat\Gamma^{cd}\hat\Gamma^{efg}\hat\Gamma^{hij})
\\
&\quad
-\tfrac{3}{32}\hat R^{ab}{}_{ef}\hat R^{cd}{}_{gh}
t_{8s}(\hat\Gamma_{ab}\hat\Gamma_{cd}\hat\Gamma^i\hat\Gamma^j)
t_{8s}(\hat\Gamma^{ef}\hat\Gamma^{gh}\nabla_i\hat{\slashed F}\nabla_j\hat{\slashed F})\,.
\end{split}
\end{equation}
Using the fact that for any $M_{1,2}$ we have, using (\ref{eq:t8s}),
\begin{equation}
\begin{split}
t_{8s}(\hat\Gamma^{ab}\hat\Gamma^{cd}M_1M_2)
&=
-\tfrac14\tr(\hat\Gamma^{abcd}M_1M_2)
-\tfrac18\tr(\hat\Gamma^{ab}M_1\hat\Gamma^{cd}M_2)
+\tfrac{1}{64}\tr(\hat\Gamma^{ab}M_1)\tr(\hat\Gamma^{cd}M_2)
\\
&\quad
+\tfrac{1}{256}\tr(\hat\Gamma^{ab}\{M_1,\hat\Gamma^e\})\tr(\{\hat\Gamma_e,M_2\}\hat\Gamma^{cd})
+(1\leftrightarrow 2)
\end{split}
\end{equation}
one finds, after some algebra,
\begin{equation}
\begin{split}
6t_{8s}t_{8s}\hat{\slashed R}^2\hat{\slashed{\mathcal R}}'^2
&=
8\big(
24B_1%
+24B_2%
-6B_4%
-48B_6
-48B_8
+24B_9%
-24B_{10}%
+24B_{11}%
\\
&\quad
-18B_{12}%
-12B_{13}%
+8B_{16}%
+8B_{17}%
-8B_{20}%
+4B_{21}%
+B_{22}
+4B_{23}
+B_{24}
\big)\,,
\end{split}
\end{equation}
where we used the basis defined by PPS, appendix \ref{app:PPS}. Using the relations (\ref{eq:Brel}), which arise by integration by parts and field redefinitions, and including the additional terms from (\ref{eq:Lhat}), which become
\begin{equation}
16\left(-12B_5+3B_{12}+6B_{14}\right)-4bB_4\,,
\end{equation}
the $\hat R^2\hat F^2$ terms reduce, upon taking $b=-48$, to
\begin{equation}
16
\left(
-24B_5
-48B_8
-24B_{10}
-6B_{12}
-12B_{13}
+12B_{14}
+8B_{16}
-4B_{20}
+B_{22}
+4B_{23}
+B_{24}
\right)\,.
\label{eq:R2dF2}
\end{equation}
This coincides with the result of PPS up to the overall coefficient, which in their case (rescaling their result by $2^{20}$ so the coefficient of $t_8t_8\hat R^4$ is $1$) is $54$. After rescaling their $\hat F\rightarrow\frac29\hat F$ this becomes $\frac83$, which differs from our result by a factor of 6. This mismatch appears to be due to missing symmetry factors in PPS \cite{Peeters:2005tb}.

Finally, these terms can be cast in a much simpler form by noting that
\begin{equation}
\begin{split}
\hat R_{abcd}&\hat R_{efgh}\tr(\hat\Gamma^c\hat\Gamma^{ef}\nabla^d\hat{\slashed F}\hat\Gamma^a\hat\Gamma^{gh}\nabla^b\hat{\slashed F})
+\hat R_{abcd}\hat R_{efgh}\tr(\hat\Gamma^e\hat\Gamma^{cd}\nabla^f\hat{\slashed F}\hat\Gamma^a\hat\Gamma^{gh}\nabla^b\hat{\slashed F})
%
\\
=&\,{}
\tfrac{64}{3}
\left(
-12B_1
-12B_2
-3B_4
+24B_6
-12B_9
+3B_{12}
-4B_{17}
+2B_{20}
-2B_{21}
\right)\,,
\end{split}
\end{equation}
which, after using the relations in (\ref{eq:Brel}), gives the following simple form for the $\hat R^2\hat F^2$ terms in the $D=11$ Lagrangian
\begin{equation}
-\tfrac32\hat R_{abcd}\hat R_{efgh}\tr(\hat\Gamma^c\hat\Gamma^{ef}\nabla^d\hat{\slashed F}\hat\Gamma^a\hat\Gamma^{gh}\nabla^b\hat{\slashed F})
-\tfrac32\hat R_{abcd}\hat R_{efgh}\tr(\hat\Gamma^e\hat\Gamma^{cd}\nabla^f\hat{\slashed F}\hat\Gamma^a\hat\Gamma^{gh}\nabla^b\hat{\slashed F})\,.
\label{eq:R2F2s}
\end{equation}

\subsection{\texorpdfstring{$\hat F^4$}{F4} terms}
Finally, we have the $\hat F^4$ terms from
\begin{equation}
\begin{split}
t_{8s}t_{8s}\hat{\slashed{\mathcal R}}'^4
&=
\tfrac{1}{6^4}t_8^{a_1a_2b_1b_2c_1c_2d_1d_2}
\nabla_{a_1}\hat F_{a_2a_3a_4a_5}\nabla_{b_1}\hat F_{b_2b_3b_4b_5}\nabla_{c_1}\hat F_{c_2c_3c_4c_5}\nabla_{d_1}\hat F_{d_2d_3d_4d_5}\\
&\quad\times t_{8s}(\hat\Gamma^{a_3a_4a_5}\hat\Gamma^{b_3b_4b_5}\hat\Gamma^{c_3c_4c_5}\hat\Gamma^{d_3d_4d_5})%
\\
&{}
+\tfrac23\nabla_{[a}\hat F_{c]a_3a_4a_5}\nabla^{[a}\hat F^{d]b_3b_4b_5}t_{8s}(\hat\Gamma^{a_3a_4a_5}\hat\Gamma_{b_3b_4b_5}\nabla^c\hat{\slashed F}\nabla_d\hat{\slashed F})
\\
&{}
-\tfrac16\nabla_{[a}\hat F_{b]a_3a_4a_5}\nabla^a\hat F^{bb_3b_4b_5}t_{8s}(\hat\Gamma^{a_3a_4a_5}\hat\Gamma_{b_3b_4b_5}\nabla_c\hat{\slashed F}\nabla^c\hat{\slashed F})
\\
&{}
+\tfrac32t_{8s}(\nabla_a\hat{\slashed F}\nabla^a\hat{\slashed F}\nabla_b\hat{\slashed F}\nabla^b\hat{\slashed F})%
\,.
\end{split}
\end{equation}
In the basis of terms defined by PPS, appendix \ref{app:PPS}, the first term becomes
\begin{equation}
\begin{split}
-36&t_8^{a_1a_2b_1b_2c_1c_2d_1d_2}\nabla_{a_1}\hat F_{a_2cd}{}^a\nabla_{b_1}\hat F_{b_2efa}\nabla_{c_1}\hat F_{c_2}{}^{b[cd}\nabla_{d_1}\hat F_{d_2b}{}^{ef]}
\\
&{}
+\tfrac23t_8^{a_1a_2b_1b_2c_1c_2d_1d_2}\nabla_{a_1}\hat F_{a_2efg}\nabla_{b_1}\hat F_{b_2}{}^{efg}\nabla_{c_1}\hat F_{c_2hij}\nabla_{d_1}\hat F_{d_2}{}^{hij}
\\
=&
-\tfrac{3}{16}C_1
+\tfrac43C_2
+\tfrac{10}{3}C_3
-2C_4
-2C_5
-6C_6
+6C_7
-18C_8
-\tfrac{10}{3}C_9
-72C_{10}
\\
&\quad
+12C_{11}
-144C_{12}
-48C_{13}
+66C_{14}
+60C_{15}
+4C_{16}
+84C_{17}
-96C_{19}
+48C_{20}
\\
&\quad
+18C_{21}
+2C_{22}
+18C_{23}
+36C_{24}\,.
\end{split}
\end{equation}
The second becomes
\begin{equation}
\begin{split}
\tfrac23&\nabla_{[a}\hat F_{c]a_3a_4a_5}\nabla^{[a}\hat F^{d]b_3b_4b_5}t_{8s}(\hat\Gamma^{a_3a_4a_5}\hat\Gamma_{b_3b_4b_5}\nabla^c\hat{\slashed F}\nabla_d\hat{\slashed F})
\\
=&
C_2
-\tfrac43C_3
-4C_4
+8C_5
-4C_6
-72C_7
+72C_8
-\tfrac43C_9
-24C_{10}
+4C_{11}
-96C_{12}
-24C_{13}
\\
&\quad
-6C_{14}
+36C_{15}
+\tfrac43C_{16}
-30C_{18}
-120C_{19}
+48C_{20}
-30C_{21}
+\tfrac83C_{22}
-90C_{23}
-24C_{24}\,.
\end{split}
\end{equation}
Similarly, the third becomes
\begin{equation}
\begin{split}
-\tfrac16&\nabla_{[a}\hat F_{b]a_3a_4a_5}\nabla^a\hat F^{bb_3b_4b_5}t_{8s}(\hat\Gamma^{a_3a_4a_5}\hat\Gamma_{b_3b_4b_5}\nabla_c\hat{\slashed F}\nabla^c\hat{\slashed F})
\\
=&
-\tfrac18C_1
+\tfrac13C_2
+\tfrac83C_3
-24C_4
+4C_5
+36C_7
-36C_8
+36C_{10}
+6C_{14}
-24C_{15}
\\
&\quad
-48C_{17}
-6C_{18}
-24C_{19}
+6C_{21}
-6C_{23}
-24C_{24}\,,
\end{split}
\end{equation}
while the fourth term becomes
\begin{equation}
\begin{split}
\tfrac32t_{8s}&(\nabla_a\hat{\slashed F}\nabla^a\hat{\slashed F}\nabla_b\hat{\slashed F}\nabla^b\hat{\slashed F})
\\
=&
-\tfrac{1}{48}C_1
+\tfrac23C_3
-6C_4
+2C_5
+\tfrac23C_6
+18C_7
-18C_8
+\tfrac23C_9
+12C_{10}
+6C_{14}
\\
&{}
-12C_{15}
-12C_{17}
+18C_{21}
-\tfrac23C_{22}
+18C_{23}\,.
\end{split}
\end{equation}
Finally, the terms coming from $\hat L_2$ in (\ref{eq:L2}) are
\begin{equation}
\begin{split}
&{}
-\tfrac{1}{12}C_1
+\tfrac83C_3
+8C_4
+8C_5
+\tfrac83C_6
-24C_7
+24C_8
-\tfrac83C_9
-144C_{10}
+24C_{14}
\\
&\quad
+144C_{15}
+144C_{17}
-24C_{21}
+8C_{22}
-24C_{23}\,.
\end{split}
\end{equation}
Putting this together we find, using the relations (\ref{eq:Crel}),
\begin{equation}
\hat L_{\hat F^4}+\sum_i d_i\tilde Z_i
=
\tfrac83
\left(
3C_5
+C_6
-9C_8
+C_9
-72C_{12}
+9C_{14}
+18C_{17}
-9C_{18}
-72C_{19}
-C_{22}
\right)\,,
\end{equation}
where we have taken the non-zero coefficients in (\ref{eq:Lhatp}) to be
\begin{equation}
\begin{split}
&
c_2=12\,,\quad
c_3=24\,,\quad
c_5=\tfrac13\,,\quad
c_7=-\tfrac{10}{3}\,,\quad
c_8=\tfrac{68}{3}\,,\quad
c_9=12
\end{split}
\end{equation}
and
\begin{equation}
\begin{split}
&d_1=-4\,,\quad
d_2=-\tfrac{20}{3}\,,\quad
d_3=\tfrac14\,,\quad
d_4=-120\,,\quad
d_5=12\,,\quad
d_6=-60\,,\quad
d_7=-\tfrac16\,,\quad
\\
&\quad
d_8=2\,,\quad
d_9=144\,.
\end{split}
\end{equation}
This agrees with the result of PPS up to the overall coefficient, which in their case (rescaling their result by $2^{20}$ so that the coefficient of $t_8t_8\hat R^4$ is $1$) is $3^7/2$. After rescaling their $\hat F\rightarrow\frac29\hat F$ this becomes $\frac83$, coinciding with our result.

Finally, these terms can be written in a nicer form by noting that
\begin{equation}
\begin{split}
\tr(\nabla_a\hat{\slashed F}&\hat\Gamma_c\nabla_b\hat{\slashed F}\hat\Gamma_d\nabla^a\hat{\slashed F}\hat\Gamma^c\nabla^b\hat{\slashed F}\hat\Gamma^d)
+2\tr(\nabla_a\hat{\slashed F}\hat\Gamma_c\nabla_b\hat{\slashed F}\nabla^a\hat{\slashed F}\hat\Gamma^c\nabla^b\hat{\slashed F})
-3\tr(\nabla_a\hat{\slashed F}\nabla_b\hat{\slashed F}\nabla^a\hat{\slashed F}\nabla^b\hat{\slashed F})
\\
=&\,{}
64\nabla_a\hat F_{cdef}\nabla^a\hat F^{cdgh}\nabla_b\hat F_{ghij}\nabla^b\hat F^{efij}
-32\nabla_a\hat F_{cdef}\nabla^a\hat F_{ghij}\nabla_b\hat F^{cdgh}\nabla^b\hat F^{efij}
\\
&\quad
-\tfrac{512}{9}\nabla_a\hat F_{cdef}\nabla_b\hat F^{gdef}\nabla^a\hat F^{chij}\nabla^b\hat F_{ghij}
-\tfrac89\nabla_a\hat F_{cdef}\nabla^a\hat F^{cdef}\nabla_b\hat F_{ghij}\nabla^b\hat F^{ghij}
\\
&\quad
+\tfrac{16}{3}\nabla_a\hat F_{cdef}\nabla_b\hat F^{cdef}\nabla^a\hat F_{ghij}\nabla^b\hat F^{ghij}
\\
=&\,{}
\tfrac{256}{9}
\left(
-\tfrac{1}{32}C_1
+3C_4
-6C_5
-2C_6
-9C_7
+9C_8
-3C_9
+9C_{21}
+9C_{23}
\right)
\,,
\end{split}
\end{equation}
which, after using the relations (\ref{eq:Crel}), gives the following simple form for the $\hat F^4$ terms in the $D=11$ Lagrangian
\begin{equation}
-\tfrac{3}{32}
\tr(\nabla_a\hat{\slashed F}\hat\Gamma_c\nabla_b\hat{\slashed F}\hat\Gamma_d\nabla^a\hat{\slashed F}\hat\Gamma^c\nabla^b\hat{\slashed F}\hat\Gamma^d)
-\tfrac{3}{16}\tr(\nabla_a\hat{\slashed F}\hat\Gamma_c\nabla_b\hat{\slashed F}\nabla^a\hat{\slashed F}\hat\Gamma^c\nabla^b\hat{\slashed F})
+\tfrac{9}{32}\tr(\nabla_a\hat{\slashed F}\nabla_b\hat{\slashed F}\nabla^a\hat{\slashed F}\nabla^b\hat{\slashed F})\,.
\label{eq:F4s}
\end{equation}

\section{Simplifying \texorpdfstring{$t_8t_8$}{t8t8} dilaton terms}\label{app:Weyl}
Here we consider the term
\begin{equation}
t_8t_8R^4
\end{equation}
in an arbitrary number of dimensions. Performing a Weyl rescaling
\begin{equation}
e^a=e^{k\Phi}e'^a
\label{eq:Weyl2}
\end{equation}
gives
\begin{equation}
R_{abcd}=e^{-2k\Phi}(R'_{abcd}+D_{abcd}+\ldots)\qquad D^{ab}{}_{cd}=4k\delta^{[a}_{[c}\nabla^{\phantom{a}}_{d]}\nabla^{b]}\Phi\,,
\end{equation}
where the higher-order terms denoted by the ellipsis will not be relevant for us. We get
\begin{equation}
t_8t_8R^4=t_8t_8R'^4+4t_8t_8DR'^3+6t_8t_8D^2R'^2+4t_8t_8D^3R'+t_8t_8D^4+\ldots\,.
\label{eq:R4Weyl}
\end{equation}
Here we will simplify the last four terms, which involve the dilaton, by using field redefinitions and integration by parts, at the linearized level, i.e. dropping total derivative terms, higher-order terms and terms proportional to the lowest-order equations of motion.

The dilaton terms can be treated all at once, since they all have the form
\begin{equation}
t_8t_8DR^{(1)}R^{(2)}R^{(3)}\,,
\end{equation}
with each $R^{(i)}$ being either $R'$ or $D$. Note that $D$ and $R'$ both have the symmetries and Bianchi identities of the Riemann tensor and, furthermore, that the divergence of both is proportional to the equations of motion plus higher-order terms. Any such terms can be removed by field redefinitions. Therefore, we can work with $R^{(i)}$ as if it were a Riemann tensor.

Integrating by parts twice, this becomes
\begin{equation}
4k\Phi t_8t_8\nabla^2(R^{(1)}R^{(2)}R^{(3)})
\end{equation}
where
\begin{equation}
\begin{split}
t_8t_8\nabla^2(R^{(1)}R^{(2)}R^{(3)})
&=
32\delta^e_{[a}\nabla^{\phantom{e}}_{b]}\nabla^f(R^{(1)bc}{}_{fg}R^{(2)}_{cd}{}^{gh}R^{(3)da}{}_{he})
+64\delta^{[e}_{[a}\nabla^{\phantom{[e}}_{b]}\nabla^{f]}(R^{(1)}_{cdfg}R^{(2)dagh}R^{(3)bc}{}_{he})
\\
&\quad
-32\delta^e_{[a}\nabla^{\phantom{e}}_{b]}\nabla^f(R^{(1)bc}{}_{ef}R^{(2)}_{cdgh}R^{(3)dagh})
-16\delta^e_{[a}\nabla^{\phantom{e}}_{b]}\nabla^f(R^{(1)bcgh}R^{(2)}_{cdef}R^{(3)da}{}_{gh})
\\
&\quad
+4\delta^e_a\nabla_b\nabla^f(R^{(1)abgh}R^{(2)}_{cdef}R^{(3)cd}{}_{gh})
+2\delta^e_a\nabla_b\nabla^f(R^{(1)ab}{}_{ef}R^{(2)}_{cdgh}R^{(3)cdgh})
\\
&\quad
+\mathrm{perm}(1,2,3)\,.
\label{eq:t8t8d2R3}
\end{split}
\end{equation}
Consider the first term,
\begin{equation}
\begin{split}
32\delta^e_{[a}&\nabla^{\phantom{e}}_{b]}\nabla^f(R^{(1)bc}{}_{fg}R^{(2)}_{cd}{}^{gh}R^{(3)da}{}_{he})
+\mathrm{perm}(1,2,3)
\\
=&
16\nabla_b\nabla_f(R^{(1)bcfg}R^{(2)}_{cdgh}R^{(3)dh})
-16\nabla_a\nabla^f(R^{(1)bc}{}_{fg}R^{(2)}_{cd}{}^{gh}R^{(3)da}{}_{hb})
+\mathrm{perm}(1,2,3)\,,
\end{split}
\end{equation}
where the second term further becomes
\begin{equation}
\begin{split}
-16&\nabla_a\nabla^f(R^{(1)bc}{}_{fg}R^{(2)}_{cd}{}^{gh}R^{(3)da}{}_{hb})
+\mathrm{perm}(1,2,3)
\\
=&
8\nabla_a(R^{(1)bc}{}_{fg}\nabla^hR^{(2)}_{cd}{}^{fg}R^{(3)da}{}_{hb})
+8\nabla_a(R^{(1)bc}{}_{fg}R^{(2)}_{cd}{}^{gh}\nabla^aR^{(3)fd}{}_{hb})
+\mathrm{perm}(1,2,3)
+\ldots\,,
\end{split}
\end{equation}
where we used the anti-symmetry in $df$ in the second term. This further simplifies to
\begin{equation}
\begin{split}
&
8\nabla_a\nabla^h(R^{(1)bc}{}_{fg}R^{(2)}_{cd}{}^{fg}R^{(3)da}{}_{hb})
+4\nabla^a(\nabla_fR^{(1)cd}{}_{gh}R^{(2)fbgh}R^{(3)}_{bacd})
\\
&\quad
-8\nabla_a(R^{(1)}_f{}^{cb}{}_gR^{(2)}_{cd}{}^{gh}\nabla^aR^{(3)df}{}_{hb})
+2\nabla_a(R^{(1)}_{cdef}R^{(2)cdgh}\nabla^aR^{(3)ef}{}_{gh})
+\mathrm{perm}(1,2,3)
+\ldots
\\
=&
8\nabla_a\nabla^h(R^{(1)bc}{}_{fg}R^{(2)}_{cd}{}^{fg}R^{(3)da}{}_{hb})
+4\nabla^a\nabla_f(R^{(1)cd}{}_{gh}R^{(2)fbgh}R^{(3)}_{bacd})
\\
&\quad
-\tfrac83\nabla^2(R^{(1)}_f{}^{cb}{}_gR^{(2)}_{cd}{}^{gh}R^{(3)df}{}_{hb})
+\tfrac43\nabla^2(R^{(1)}_{cdef}R^{(2)cdgh}R^{(3)ef}{}_{gh})
+\mathrm{perm}(1,2,3)
+\ldots\,.
\end{split}
\end{equation}
The first term in (\ref{eq:t8t8d2R3}) therefore reduces to
\begin{equation}
\begin{split}
32\delta^e_{[a}\nabla^{\phantom{e}}_{b]}\nabla^f&(R^{(1)bc}{}_{fg}R^{(2)}_{cd}{}^{gh}R^{(3)da}{}_{he})
+\mathrm{perm}(1,2,3)
\\
=&
16\nabla_b\nabla_f(R^{(1)bcfg}R^{(2)}_{cdgh}R^{(3)dh})
+8\nabla_a\nabla^h(R^{(1)bc}{}_{fg}R^{(2)}_{cd}{}^{fg}R^{(3)da}{}_{hb})
\\
&\quad
+4\nabla^a\nabla_f(R^{(1)cd}{}_{gh}R^{(2)fbgh}R^{(3)}_{bacd})
-\tfrac83\nabla^2(R^{(1)}_f{}^{cb}{}_gR^{(2)}_{cd}{}^{gh}R^{(3)df}{}_{hb})
\\
&\quad
+\tfrac43\nabla^2(R^{(1)}_{cdef}R^{(2)cdgh}R^{(3)ef}{}_{gh})
+\mathrm{perm}(1,2,3)
+\ldots\,.
\end{split}
\end{equation}
Next we consider the second term in (\ref{eq:t8t8d2R3}). It becomes
\begin{equation}
\begin{split}
64\delta^{[e}_{[a}\nabla^{\phantom{[e}}_{b]}\nabla^{f]}&(R^{(1)}_{cdfg}R^{(2)dagh}R^{(3)bc}{}_{he})
\\
=&
16\nabla_b\nabla^f(R^{(1)}_{cdfg}R^{(2)dagh}R^{(3)bc}{}_{ha})
-16\nabla_b\nabla^e(R^{(1)}_{cdag}R^{(2)dagh}R^{(3)bc}{}_{he})
\\
&\quad
+16\nabla_a\nabla^e(R^{(1)}_{cdbg}R^{(2)dagh}R^{(3)bc}{}_{he})
+16\nabla_a\nabla^f(R^{(1)}_{cdfg}R^{(2)dagh}R^{(3)c}{}_h)
\\
=&
8\nabla_b\nabla^e(R^{(1)}_{cdag}R^{(2)dhag}R^{(3)bc}{}_{he})
+4\nabla_b\nabla^f(R^{(1)}_{cfdg}R^{(2)dgah}R^{(3)bc}{}_{ha})
\\
&\quad
-4\nabla_a\nabla^e(R^{(1)}_{cdgh}R^{(2)ghab}R^{(3)cd}{}_{be})
+16\nabla_a\nabla^f(R^{(1)}_{cdfg}R^{(2)dagh}R^{(3)c}{}_h)\,.
\end{split}
\end{equation}
The first adds to the second term of the previous expression and the next two add to the third term. Using this in (\ref{eq:t8t8d2R3}) many terms cancel and we are left with
\begin{equation}
\begin{split}
t_8t_8\nabla^2(R^{(1)}R^{(2)}R^{(3)})
&=
16\nabla_b\nabla_f(R^{(1)dh}R^{(2)bcfg}R^{(3)}_{cdgh})
+16\nabla_b\nabla^f(R^{(1)d}{}_hR^{(2)}_{cdfg}R^{(3)bcgh})
\\
&\quad
+16\nabla_b\nabla_f(R^{(1)cf}R^{(2)}_{cdgh}R^{(3)dbgh})
+2\nabla^b\nabla^f(R^{(1)}_{bf}R^{(2)}_{cdgh}R^{(3)cdgh})
\\
&\quad
-\tfrac83\nabla^2(R^{(1)}_f{}^{cb}{}_gR^{(2)}_{cd}{}^{gh}R^{(3)df}{}_{hb})
+\tfrac43\nabla^2(R^{(1)}_{cdef}R^{(2)cdgh}R^{(3)ef}{}_{gh})
\\
&\quad
+\mathrm{perm}(1,2,3)
+\ldots\,.
\end{split}
\end{equation}

The dilaton terms generated by the Weyl transformation can be read off from
\begin{equation}
\begin{split}
t_8t_8DR^{(1)}R^{(2)}R^{(3)}
=&
4k\Phi t_8t_8\nabla^2(R^{(1)}R^{(2)}R^{(3)})
+\ldots
\\
=&
8k\nabla^a\nabla^b\Phi
\big[
8R^{(1)cd}R^{(2)}_{aghb}R^{(3)cghd}
+8R^{(1)}_{cd}R^{(2)}_{ag}{}^{hc}R^{(3)}_{bh}{}^{gd}
\\
&\quad
-8R^{(1)}_{ac}R^{(2)cdgh}R^{(3)}_{bdgh}
+R^{(1)}_{ab}R^{(2)}_{cdgh}R^{(3)cdgh}
+\mathrm{perm}(1,2,3)
\big]
+\ldots\,,
\end{split}
\end{equation}
where we have integrated twice by parts and used the fact that $\nabla^2\Phi$ vanishes by the linearized equations of motion. These terms are at least quadratic in the dilaton, since $R^{(1)}_{cd}\propto\nabla_c\nabla_d\Phi$ regardless of whether $R^{(1)}$ is $R'$ or $D$.

We can further simplify terms where at least one of $R^{(2,3)}$ is $D$. Such terms have the form
\begin{equation}
\begin{split}
8k&\nabla^a\nabla^b\Phi
\left(
8R^{(1)cd}D_{aghb}R^{(3)cghd}
+8R^{(1)}_{cd}D_{ag}{}^{hc}R^{(3)}_{bh}{}^{gd}
-8R^{(1)}_{ac}D^{cdgh}R^{(3)}_{bdgh}
+R^{(1)}_{ab}D_{cdgh}R^{(3)cdgh}
\right)
\\
=&
32k^2\nabla^a\nabla^b\Phi
\left(
6R^{(1)cd}\nabla_a\nabla^e\Phi R^{(3)}_{bcde}
+3R^{(1)}_{ab}\nabla^c\nabla^d\Phi R^{(3)}_{cd}
-2R^{(1)cd}\nabla_a\nabla_c\Phi R^{(3)}_{bd}
\right)
+\ldots
\\
=&
32k^2\nabla^a\nabla^b\Phi
\left(
3R^{(1)}_{ab}\nabla^c\nabla^d\Phi R^{(3)}_{cd}
-2R^{(1)cd}\nabla_a\nabla_c\Phi R^{(3)}_{bd}
\right)
+\ldots\,,
\end{split}
\label{eq:D3terms}
\end{equation}
where we noted that the first term in the second line can be removed by integrations by parts and field redefinitions.\footnote{Indeed, this term is proportional to
\begin{equation*}
\begin{split}
\nabla^a\nabla^b\Phi\nabla^c\nabla^d\Phi\nabla_a\nabla^e\Phi R^{(3)}_{bcde}
=&
-\nabla^d\nabla^a\nabla^b\Phi\nabla^c\Phi\nabla_a\nabla^e\Phi R^{(3)}_{bcde}
+\ldots
\\
=&
-\tfrac12\nabla^d\nabla^a\nabla^b\Phi\nabla_a(\nabla^c\Phi\nabla^e\Phi)R^{(3)}_{bcde}
+\ldots
\\
=&
-\tfrac14\nabla^2(\nabla^d\nabla^b\Phi\nabla^c\Phi\nabla^e\Phi)R^{(3)}_{bcde}
+\tfrac14\nabla^d\nabla^b\Phi\nabla^2(\nabla^c\Phi\nabla^e\Phi)R^{(3)}_{bcde}
+\ldots
\\
=&
-\tfrac12\nabla^d\nabla^c\Phi\nabla^a\nabla^b\Phi\nabla_a\nabla^e\Phi R^{(3)}_{bcde}
+\ldots\,,
\end{split}
\end{equation*}
which is the same as the term we started with up to the coefficient, i.e. this term vanishes.}

Using these results, we can find the dilaton terms generated by the Weyl transformation (\ref{eq:R4Weyl}). Noting that the metric equation of motion (before the Weyl transformation) is
\begin{equation}
R_{ef}=R'_{ef}+D_{ef}=l\nabla_e\nabla_f\Phi+\ldots\,,
\end{equation}
where the constant $l$ depends on the theory we start with and
\begin{equation}
D_{ef}=k(D-2)\nabla_e\nabla_f\Phi+\ldots\,,
\end{equation}
we finally find that under the Weyl rescaling in (\ref{eq:Weyl2}) we have
\begin{equation}
t_8t_8R^4
=
t_8t_8R'^4
+96k(2l-k(D-2))\nabla^a\nabla^b\Phi\nabla^c\nabla^d\Phi X_{abcd}
+\ldots
\end{equation}
with
\begin{equation}
\begin{split}
X_{ab}{}^{cd}
=&
8R'_{aghb}R'^{cghd}
+8R'_{ag}{}^{hc}R'_{bh}{}^{gd}
-8\delta^c_aR'^{dfgh}R'_{bfgh}
+\delta^c_a\delta^d_bR'_{efgh}R'^{efgh}
\\
&\quad
+2k(2l-k(D-2))\big[3\nabla_a\nabla_b\Phi\nabla^c\nabla^d\Phi
-2\nabla_a\nabla^c\Phi\nabla_b\nabla^d\Phi\big]\,.
\end{split}
\end{equation}

\bibliographystyle{JHEP}
\bibliography{refs.bib}
\end{document}